\documentclass[conference]{IEEEtran}
\IEEEoverridecommandlockouts
\usepackage[cmex10]{amsmath}

\newtheorem{example}{Example}
\newtheorem{definition}{Definition} 
\newtheorem{proposition}{Proposition}
\newtheorem{lemma}{Lemma}
\newtheorem{corollary}{Corollary}
\newtheorem{remark}{Remark}

\newenvironment{snotation}{\indent\ti{Notation:}}{}

\newenvironment{mmproof}{\hspace{8pt}\ti{Proof:}}{}

\newcommand{\pnt}[1]{{\mbox{\boldmath $#1$}}}
\newcommand{\cof}[2]{\mbox{$#1_{\boldsymbol{#2}}$}}

\newcommand{\V}[1]{\mbox{$\mathit{Vars}(#1)$}}
\newcommand{\Va}[1]{\mbox{$\mathit{Vars}(\boldsymbol{#1})$}}
\newcommand{\Vaa}[1]{\mathit{Vars}(\boldsymbol{#1})}

\newcommand{\s}[1]{\mbox{$\{#1\}$}}

\newcommand{\nGz}[2]{$G_{non-\{z\}}$}

\newcommand{\Dis}[2]{\mbox{$\mathit{Dis}(#1,#2)$}}
\newcommand{\DDis}[3]{\mbox{\Dis{\cof{#1}{#2}}{#3}}}

\newcommand{\prr}[1]{\mi{Prev}(\boldsymbol{q})}
\newcommand{\la}[1]{\mbox{$\mathit{Last}$(\pnt{#1})}}

\newcommand{\mi}[1]{\mathit{#1}}
\newcommand{\ti}[1]{\textit{#1}}
\newcommand{\tb}[1]{\textbf{#1}}

\newcommand{\meqq}[1]{$F^*_{#1}(Y_{#1}) \equiv \exists{X_{#1}}.F_{#1}(X_{#1},Y_{#1})$}
\newcommand{\Res}[3]{\mbox{$\mathit{Res}$(\pnt{#1},\pnt{#2},$#3$)}}
\newcommand{\Ds}[2]{\mbox{\pnt{#1}~$\rightarrow$ \s{#2}}}
\newcommand{\Dds}[2]{\mbox{\pnt{#1}~$\rightarrow #2$}~}
\newcommand{\DDs}[4]{\mbox{$(#1,#2,\pnt{#3})~\rightarrow$ \s{#4}}}
\newcommand{\Dss}[4]{\mbox{$(#1,#2,\pnt{#3})~\rightarrow #4$}}
\newcommand{\Xm}{\mbox{$X_{\mi{mon}}$}}
\newcommand{\Xr}{\mbox{$X_{\mi{red}}$}}
\newcommand{\ADS}{\mbox{$\mi{D_\mi{seq}^{act}}$}~} 
\newcommand{\IDS}{\mbox{$\mi{D_\mi{seq}^{inact}}$}~}
\newcommand{\ttt}{\>\>\>}

\newcommand{\Tt}{\>\>}
\newcommand{\Su}[2]{\mbox{$#1_\mi{#2}$}}
\newcommand{\Di}{\ti{DDS\_impl}~}
\newcommand{\Fcurr}{\DDis{F}{q}{\Su{X}{red}}~}
\newcommand{\Ss}[1]{\scriptsize{#1}}

\begin{document}

\title{Removal of Quantifiers by Elimination of Boundary Points}

\author{\IEEEauthorblockN{Eugene Goldberg and Panagiotis Manolios} \\
\IEEEauthorblockA{College of Computer and Information Science\\
Northeastern University\\
360 Huntington Ave., Boston MA 02115, USA \\
Email: \{eigold,pete\}@ccs.neu.edu}}

\maketitle

\begin{abstract}
We consider the problem of elimination of existential quantifiers from a Boolean CNF formula.
Our approach is based on the following observation. One can get rid of
dependency on a set of variables of a quantified CNF formula $F$
by adding resolvent clauses of $F$  eliminating boundary points. This
approach is similar to the method of quantifier elimination described in~\cite{DDS_arxiv}.
The difference of the method described in the present paper is twofold:
\begin{itemize}
\item branching is performed only on quantified variables,
\item an explicit search for boundary points is performed by calls to a  SAT-solver
\end{itemize}
Although we published the paper~\cite{DDS_arxiv} before this one, chronologically
the method  of the present report  was developed first. Preliminary presentations of
this method  were made in~\cite{DDS_SRC_2011,DDS_SRC_report}. We postponed a publication of this method due to
preparation of a patent application~\cite{DDS_patent}.
\end{abstract}

\section{Introduction}
In this paper, we are concerned with the problem of elimination of  quantified variables from a Boolean CNF formula.
 (Since we consider
only existential quantifiers, further on we omit the word ``existential''.)
Namely, we solve the following problem: given a Boolean CNF formula $\exists{X}.F(X,Y)$, find a
 Boolean CNF formula $F^*(Y)$  such
that $F^*(Y) \equiv \exists{X}.F(X,Y)$. We will refer to this problem as QEP (Quantifier Elimination Problem).
Since QEP is to find  a formula, it is not a decision problem as opposed to the problem  of solving a Quantified
Boolean Formula (QBF).  QEP occurs in numerous areas  of hardware/software design  and verification, 
 model checking~\cite{mc,mc_thesis}  being one of the most prominent applications of QEP.

A straightforward method of solving QEP for CNF formula  $\exists{X}.F(X,Y)$ is to
eliminate the variables of $X$ one by one, in the way it is done
in the DP procedure~\cite{dp}. To delete a variable $x_i$ of $X$, the DP procedure produces all possible resolvents 
on variable $x_i$ and adds them to $F$.
An obvious drawback of such a  method is that it  generates a prohibitively
large number of clauses. Another set of  QEP-solvers employ the idea of enumerating
satisfying assignments of formula $F(X,Y)$. Here is how a typical method of this kind works.
First, a CNF formula $F^+(Y)$
is built such that each clause $C$ of $F^+$ (called a blocking clause~\cite{blocking_clause}) eliminates a set
 of assignments satisfying $F(X,Y)$. By negating  $F^+(Y)$ one obtains a CNF formula $F^*(Y)$ that is a solution to
QEP.

Unfortunately, $F^+$ may be exponentially larger than $F^*$. This occurs, for instance, when 
$F(X,Y) = F_1(X_1,Y_1) \wedge \ldots \wedge F_k(X_k,Y_k)$ and  
$(X_i \cup Y_i) \cap (X_j \cup Y_j) = \emptyset$, $i \neq j$ that is when $F$ is the conjunction
of independent CNF formulas $F_i$. In this case, one can build $F^*(Y)$
 as $F^*_1 \wedge \ldots \wedge F^*_k$, where \meqq{i},$i=1,\ldots,k$. So the size of $F^*$ is linear
in  $k$  whereas  that of $F^+$ is \ti{exponential} in $k$.
 This fact implies that 
QEP-solvers based on enumeration of satisfying assignments \ti{are not compositional}.  (We say that a QEP-solver
is compositional if it reduces the problem of finding $F^*(Y)$ to $k$ independent subproblems of finding 
$F^*_i(Y_i)$,$i=1,\ldots,k$.)
Note that in practical applications, it is very important for a QEP-solver to be compositional. 
Even if $F$ does not break down into 
independent subformulas, there may be numerous branches of the search tree where such subformulas appear. 

Both kinds of QEP-solvers mentioned above have the same drawback.
A resolution-based QEP-solver can only efficiently check if a clause $C$ of  $F^*(Y)$ is correct i.e. whether it is implied by $F(X,Y)$.
But how does one know if $F^*$ contains a \ti{sufficient} set of correct clauses i.e. whether every assignment \pnt{y}
satisfying $F^*$ can be extended to (\pnt{x},\pnt{y}) satisfying $F$? A \ti{non-deterministic}  algorithm 
does not have to answer this question. Once a sufficient set of clauses  is
 derived, an oracle stops this algorithm. But a \ti{deterministic} algorithm has no oracle  and so has 
to decide for itself
when it is the right time to terminate.
One way to guarantee the correctness of termination  is to enumerate the satisfying 
assignments of $F$. The problem here is that then, the size of a deterministic derivation of $F^*$ may be exponentially
larger than that of a non-deterministic one. (Non-compositionality of QEP-solvers based
on enumeration of satisfying assignments is just a special case of this problem.)

In this paper, we introduce a new termination condition for QEP that is based on the notion of boundary points. 
A complete assignment \pnt{p} falsifying $F(X,Y)$ is an $X'$-boundary point where $X' \subseteq X$ 
if a) every clause of $F$ falsified by \pnt{p} has a variable of $X'$ and b) first condition breaks for every proper subset of $X'$.
 An $X'$-boundary point \pnt{p} is called removable
if no satisfying assignment of $F$ can be obtained from \pnt{p} by changing values of variables of $X$.
One can eliminate a removable $X'$-boundary
point by adding to $F$ a clause $C$ that is implied by $F$ and does not have a variable of $X'$.
If for a set of  variables $X''$ where $X'' \subseteq X$, formula $F(X,Y)$ does not have a removable $X'$-boundary point where $X' \subseteq X''$,
the  variables of $X''$ are \ti{redundant} in formula $\exists{X}.F(X,Y)$.  This means that  every clause with a variable of $X''$
can be removed from $F(X,Y)$.
 QEP-solving  terminates when the current formula $F(X,Y)$ (consisting of the initial clauses and resolvents) has no removable boundary points.
 A solution $F^*(Y)$ to QEP is formed  from $F(X,Y)$ by discarding every clause that has a variable of $X$.

The new termination condition allows one to address drawbacks of  the QEP-solvers mentioned above.
In contrast to the DP procedure,
\ti{only} resolvents eliminating a boundary point need to be added. This  dramatically reduces 
the number of resolvents one has to  generate.
On the other hand, a solution $F^*$ can be derived \ti{directly} without enumerating satisfying 
assignments of $F$. In particular, using the new
termination condition   makes a QEP-solver compositional.

To record the fact that all boundary removable points have been removed from a subspace of the search space,
 we introduce the notion of a dependency sequent
(D-sequent for short). Given a CNF formula $F(X,Y)$, a D-sequent has the form
 \Dss{F}{X'}{q}{X''} where \pnt{q} is a partial assignment to variables of $X$, $X' \subseteq X$, $X'' \subseteq X$.
Let \cof{F}{q} denote formula $F$  after assignments  \pnt{q} are made.
We say that the  D-sequent above holds  if 
\begin{itemize}
\item  the variables of $X'$ are redundant in \cof{F}{q},
\item the variables of $X''$ are redundant in the formula obtained from \cof{F}{q} by discarding every clause containing
a variable of $X'$.
\end{itemize}
The fact that the variables of $X'$ (respectively $X''$) are redundant in $F$ means that $F$ has no removable  $X^*$-boundary point 
where $X^* \subseteq X'$ (respectively $X^* \subseteq X''$).
 The reason for using name  ``D-sequent'' is that 
the validity of   \Dss{F}{X'}{q}{X''}  suggests
interdependency of variables of \pnt{q}, $X'$ and $X''$.

In a sense, the notion of a D-sequent generalizes that of an implicate of
formula $F(X,Y)$. Suppose, for instance, that  $F \rightarrow C$ where  $C= x_1 \vee x_2$, $x_1 \in X$, $x_2 \in X$.
After adding  $C$ to $F$, the D-sequent \Dss{F}{\emptyset}{q}{X'} where \pnt{q}=$(x_1=0,~x_2~\!=~\!0)$, 
$X' = X \setminus \{x_1,x_2\}$ becomes true.
 (An assignment falsifying $C$ makes the unassigned variables of $F$ redundant.)  But the opposite is not true. 
The D-sequent above may hold even if $F \rightarrow C$ does not. (The latter means that
\pnt{q} can be extended to an assignment satisfying $F$).

We will refer to the method of QEP-solving based on elimination of boundary points as DDS (Derivation of D-Sequents).
We will refer to the QEP-solver based on the DDS method we describe in this paper as \Di (DDS implementation).
 To reflect the progress
in elimination of boundary points of $F$, \Di uses resolution of D-sequents.
 Suppose D-sequents \DDs{F}{\emptyset}{q_1}{x_{10}} and \DDs{F}{\emptyset}{q_2}{x_{10}} have been derived
where \pnt{q_1}=($x_1\!\!=\!\!0$, $x_3=0$) 
and \pnt{q_2}=($x_1\!=\!1$, $x_4\!=\!0$).  Then a new D-sequent \DDs{F}{\emptyset}{q}{x_{10}} where 
 $\pnt{q}=(x_3\!=\!0,x_4\!=\!0)$
 can be produced from them  by resolution on variable $x_1$. 
\Di terminates as soon as D-sequent  $(F,\emptyset,\emptyset) \rightarrow X$ is 
derived, which means that the variables of $X$ are redundant in $F$ (because  every removable $X'$-boundary point where $X' \subseteq X$
has been eliminated from $F$ due to adding resolvent-clauses).
 
Our contribution is threefold.
First, we formulate a new method of quantifier elimination based on the notion 
of $X$-removable boundary points which are a generalization of  
those introduced in~\cite{bnd_pnts}. One of the advantages of this method is that it uses 
a new termination condition.
Second, we introduce the  notion of D-sequents and the operation of resolution of D-sequents. The calculus of D-sequents
is meant for building QEP-solvers based on the semantics of  boundary point elimination.
Third, we describe a QEP-solver called \Di and prove its compositionality. We show that
in contrast to a BDD-based QEP-solver that is compositional only for particular variable orderings,
\Di is compositional regardless of how branching variables are chosen.
We give preliminary experimental results that show the promise of DDS.

This paper is structured as follows. In Section~\ref{sec:basic}, we define the notions related to boundary points.
The relation between boundary points and QEP is discussed in Section~\ref{sec:red_vars}. 
 Section~\ref{sec:add_remove_clauses} describes how adding/removing clauses affects the set of
 boundary points of a formula.
 D-sequents are introduced in Section~\ref{sec:d_sequents}.
Section~\ref{sec:dds_impl} describes \ti{DDS\_impl}. The compositionality of \Di is discussed in Section~\ref{sec:compos}.
Section~\ref{sec:experiments} describes  experimental results. Some background in given in Section~\ref{sec:background}.
Section~\ref{sec:conclusion} summarizes this paper.

\section{Basic Definitions}
\label{sec:basic}
%
%
\begin{snotation}
Let $F$ be a CNF formula and $C$ be a clause. We denote by \V{F} (respectively \V{C}) the set of variables of $F$
(respectively of $C$). If \pnt{q} is a partial assignment to \V{F},  \Va{q} denotes the 
variables assigned in \pnt{q}.
\end{snotation}\\
%
%
\begin{snotation}
In this paper, we consider a quantified CNF formula  $\exists{X}.F(X,Y)$ 
where   $X \cup Y = $ \V{F} and $X \cap Y = \emptyset$.
\end{snotation}
%
%
\begin{definition}
\label{def:QEP}
A CNF formula $F^*(Y)$ is a solution to the Quantifier Elimination Problem (\textbf{QEP})
if $F^*(Y) \equiv \exists{X}.F(X,Y)$. 
\end{definition}
%
%
\begin{definition}
Given a CNF formula $G(Z)$, a complete assignment to the variables of $Z$ is  called \textbf{a point}.
\end{definition}
%
%
\begin{definition}
\label{def:X_clauses}
Let $G(Z)$ be a CNF formula and $Z' \subseteq Z$. A clause $C$ of $G$ is called a \pnt{Z'}\textbf{-clause} if
\V{C} $\cap~Z'~\neq \emptyset$. Otherwise, $C$ is called a \textbf{non-}\pnt{Z'}\textbf{-clause}.
\end{definition}
 %
%
\begin{definition}  
\label{def:bnd_pnt}
Let $G(Z)$ be a CNF formula and $Z' \subseteq Z$. A point  \pnt{p} is called a  \pnt{Z'}\textbf{-boundary point} of $G$
if  $G(\pnt{p})= 0$ and
\begin{enumerate}
\item Every clause of $G$ falsified by \pnt{p} is a $Z'$-clause. 
\label{enum:item2}
\item  Condition~\ref{enum:item2} breaks for every proper  subset of $Z'$.
\end{enumerate}
\end{definition}
A $Z'$-boundary point \pnt{p} is at least $|Z'|$ flips away from  a point \pnt{p^*}, $G(\pnt{p^*})=1$
(if \pnt{p^*} exists and only variables of $Z'$ are allowed to be changed), hence the name ``boundary''.

Let \pnt{p} be a $Z'$-boundary point of $G(Z)$ where $Z'=\s{z}$. Then every clause of $G$ falsified by \pnt{p}
contains variable $z$. This special class of boundary points  was introduced in \cite{date02,bnd_pnts}.

%
%
\begin{definition}
\label{def:rem_bnd_pnt}
Point \pnt{p} is called a \pnt{Z'}\textbf{-removable boundary point} of $G(Z)$ where $Z' \subseteq Z$ if 
\pnt{p} is a $Z''$-boundary point where $Z'' \subseteq Z'$ and  there is a clause $C$  such that 
\begin{itemize}
\item \pnt{p} falsifies $C$;
\item $C$ is a non-$Z'$-clause; 
\item $C$ is implied by   the conjunction of $Z'$-clauses of $G$.
\end{itemize}
Adding clause $C$ to $G$ eliminates \pnt{p} as a $Z''$-boundary point (\pnt{p} falsifies  clause $C$
and $C$ has  no variables of $Z''$).
\end{definition}
%
%
\begin{proposition} 
\label{prop:remv_bpnt}
Point \pnt{p} is a $Z'$-removable boundary point of a CNF formula $G(Z)$
iff no point \pnt{p^*} obtained from \pnt{p} by changing values of (some) variables of $Z'$ satisfies $G$.
\end{proposition}

The proofs are given in the Appendix.
\begin{example}
Let CNF formula $G$ consist of four clauses: 
$C_1=z_1 \vee z_2$, $C_2=z_3 \vee z_4$, $C_3=\overline{z_1} \vee z_5$, $C_4=\overline{z_3} \vee z_5$.
Let \pnt{p}=$(z_1\!=\!0,z_2\!=\!0,z_3\!=\!0,z_4\!=\!0,z_5\!=\!0)$. Point \pnt{p} falsifies only $C_1$ and $C_2$.
Since both $C_1$ and $C_2$ contain a variable of $Z''=\{z_1,z_3\}$, \pnt{p} is a $Z''$-boundary point. 
(Note that \pnt{p} is also, for instance, 
a $\{z_2,z_4\}$-boundary point.) Let us check if point \pnt{p} is  a $Z'$-removable boundary point
 where $Z'=\{z_1,z_3,z_5\}$. One condition of Definition~\ref{def:rem_bnd_pnt} is met:
\pnt{p} is a  $Z''$-boundary point, $Z'' \subseteq Z'$. However, 
the point \pnt{p^*} obtained from \pnt{p} by flipping the values of $z_1$,$z_3$,$z_5$ satisfies $G$.
So, according to Proposition~\ref{prop:remv_bpnt}, \pnt{p} is not a $Z'$-removable boundary point
(i.e. the clause $C$  of Definition~\ref{def:rem_bnd_pnt} does not exist for \pnt{p}).
\end{example} 
\begin{definition}
We will say that a boundary point \pnt{p} of $F(X,Y)$ is just \tb{removable} if it is $X$-removable.
\end{definition}
\begin{remark}
\label{rem:bnd_pnt_removability}
Informally, a boundary point \pnt{p} of $F(X,Y)$ is removable only if there exists a clause $C$ implied by $F$ and falsified by \pnt{p} such that
$\V{C} \subseteq Y$.  The fact that an $X''$-boundary point \pnt{p} is not $X'$-removable (where $X'' \subseteq X'$)
also means that \pnt{p} is not removable. The opposite is not true.
\end{remark}

\section{$X$-Boundary Points and Quantifier Elimination}
\label{sec:red_vars}
In this section, we relate QEP-solving and boundary points. First we define the notion of redundant variables
in the context of boundary point elimination (Definition~\ref{def:glob_red_vars}). Then we show
that monotone variables are redundant (Proposition~\ref{prop:mon_var_red}). Then we prove that
clauses containing variables of $X'$, $X' \subseteq X$ can be removed from formula $\exists{X}.F(X,Y)$ 
if and only if the variables of $X'$ are redundant in $F$
(Proposition~\ref{prop:red_vars}).

%
%
\begin{definition}
\label{def:glob_red_vars}
Let $F(X,Y)$ be a CNF formula and  $X' \subseteq X$.
We will say that the variables of $X'$ are  \textbf{redundant} in $F$ 
 if $F$ has no removable $X''$-boundary point where $X'' \subseteq X$.  
\end{definition}
%
%
\begin{proposition}
\label{prop:mon_var_red}
  Let $G(Z)$ be a CNF formula and $z$ be a monotone variable of $F$. (That is clauses of $G$ contain the literal of $z$
 of only one polarity.)  Then  $z$ is redundant in $G$.
\end{proposition}
%
%
\begin{definition}
\label{def:discard}
Let $F(X,Y)$ be a CNF formula. Denote by {\boldmath \Dis{F}{X'}}  where $X' \subseteq X$ the CNF formula obtained
from $F(X,Y)$ by discarding all $X'$-clauses. 
\end{definition}
%
%
\begin{proposition}
\label{prop:red_vars}
Let  $F(X,Y)$ be a CNF formula and $X'$ be a subset of  $X$.
Then $\exists{X}.F(X,Y) \equiv \exists{(X \setminus X')}.\Dis{F}{X'}$ iff the variables of $X'$ are redundant
in $F$.
\end{proposition}
%
%
\begin{corollary} 
\label{corol:red_vars}
Let $F(X,Y)$ be a CNF formula. Let   $F^*(Y)$~=~\Dis{F}{X}. 
Then $F^*(Y) \equiv \exists{X}.F(X,Y)$  holds iff the variables of $X$ are redundant in $F$.
\end{corollary}

\section{Appearance of Boundary Points When Adding/Removing Clauses}
\label{sec:add_remove_clauses}
In this section, we give two theorems later used in Proposition~\ref{prop:d_seq_derivation} (about 
D-sequents  built by \ti{DDS\_impl}).
They  describe the type of clauses one can  add to (or remove from) $G(Z)$ without creating 
 a new  \s{z}-removable boundary point where $z \in Z$.
\begin{proposition}
\label{prop:adding_clauses}
Let $G(Z)$ be a CNF formula. Let $G$ have no \s{z}-removable boundary points.
Let $C$ be a clause. Then the formula $G \wedge C$ does not have a \s{z}-removable boundary point 
if at least one of the following conditions hold: \linebreak a) $C$ is implied by $G$; b) $z \notin$ \V{C}.
\end{proposition}

\begin{proposition}
\label{prop:removing_clauses}
Let $G(Z)$ be a CNF formula. Let $G$ have no \s{z}-removable boundary points.
Let $C$ be a \s{z}-clause of $G$. Then the CNF  formula $G'$ where $G' = G \setminus \{C\}$ does not have a
\s{z}-removable boundary point.
\end{proposition}

\begin{remark}
\label{rem:add_rem_clauses}
According to Propositions~\ref{prop:adding_clauses} and ~\ref{prop:removing_clauses}, adding  
clause $C$ to a CNF formula $G$ or removing $C$ from $G$ may produce
a new \s{z}-removable boundary point only if:
\begin{itemize}
\item  one adds to $G$  a \s{z}-clause $C$ that is not implied by $G$~or 
\item  one removes from $G$ a clause $C$ that is not a \s{z}-clause.
\end{itemize}
\end{remark}

\section{Dependency Sequents (D-sequents)}
\label{sec:d_sequents}
%
%
\subsection{General Definitions and Properties}
In this subsection, we introduce D-sequents (Definition~\ref{def:d_sequent}) and  resolution of D-sequents
(Definition~\ref{def:res_rule}). Proposition~\ref{prop:form_replacement}
states that a D-sequent remains true if  resolvent-clauses are added to $F$. 
 The soundness of resolving
D-sequents is shown in Proposition~\ref{prop:res_rule}.
%
%
\begin{definition}
\label{def:cofactor}
Let $F$ be a CNF formula and \pnt{q} be a partial assignment to \V{F}. 
Denote by \cof{F}{q} the CNF formula obtained from $F$ by
\begin{itemize}
\item removing the literals of (unsatisfied) clauses of $F$ that are set to 0 by \pnt{q},
\item removing the clauses of $F$ satisfied by \pnt{q},
\end{itemize}
\end{definition}
%
%
\begin{definition}
\label{def:d_sequent}
Let $F(X,Y)$ be a CNF formula. Let \pnt{q} be a partial assignment to variables of $X$
and $X'$ and $X''$ be  subsets of $X$ such that \Va{q}, $X'$, $X''$ do not overlap.
A dependency sequent (\textbf{D-sequent}) $S$ has the form
$(F,X',\pnt{q}) \rightarrow X''$. We will say that $S$ holds  if
\begin{itemize}
\item the variables of  $X'$ are redundant  in \cof{F}{q} (see Definition~\ref{def:cofactor}),
\item the variables of $X''$ are redundant in \DDis{F}{q}{X'} (see Definition~\ref{def:discard}).
\end{itemize}
\end{definition}
%
%
\begin{example}
Let CNF formula $F(X,Y)$ where $X =\{x_1,x_2\}$, $Y=\{y_1,y_2\}$
consist of two clauses: $C_1=x_1 \vee y_1$ and $C_2=\overline{x}_1 \vee x_2 \vee y_2$.
Note that variable $x_2$ is monotone and hence redundant in $F$ (due to Proposition~\ref{prop:mon_var_red}).
After discarding the clause $C_2$ (containing the redundant variable $x_2$), variable $x_1$ becomes redundant.
Hence, the D-sequent $(F,\s{x_2},\emptyset) \rightarrow \s{x_1}$ holds.

\end{example}
%
%
\begin{proposition}
\label{prop:form_replacement}
Let $F^+(X,Y)$ be a CNF formula obtained from $F(X,Y)$ by adding some resolvents of clauses of $F$.
Let \pnt{q} be a partial assignment to variables of $X$ and $X' \subseteq X$.
Then the fact that D-sequent \Dss{F}{X'}{q}{X''} holds implies that
 \Dss{F^+}{X'}{q}{X''}   holds too. The opposite is not true.
\end{proposition}
%
%
\begin{definition}
\label{def:res_part_assgns}
Let $F(X,Y)$ be a CNF formula and \pnt{q'}, \pnt{q''} be partial assignments to $X$.
Let $\Va{q'}~\cap \Va{q''}$ contain exactly one variable $x$ for which \pnt{q'} and \pnt{q''}
have the opposite values. Then the partial assignment \pnt{q} such that
\begin{itemize}
\item \Va{q} = ($(\Va{q'}~\cup \Va{q''}) \setminus \s{x}$,
\item the value of each variable $x^*$ of \Va{q} is equal to that of $x^*$ in
$\Va{q'} \cup \Va{q''}$. 
\end{itemize}
is  denoted as \Res{q'}{q''}{x}  and called \textbf{the resolvent of} \pnt{q'},\pnt{q''} on $x$.
Assignments \pnt{q'} and \pnt{q''} are called resolvable on $x$. 
\end{definition}
%
%
\begin{proposition}
\label{prop:res_rule}
Let $F(X,Y)$ be a CNF formula. Let D-sequents $S_1$ and $S_2$ be equal to  \Dss{F}{X_1}{q_1}{X'} and
\Dss{F}{X_2}{q_2}{X'} respectively. Let \pnt{q_1} and \pnt{q_2}
be resolvable on variable $x$.  Denote by \pnt{q} the partial assignment \Res{q_1}{q_2}{x}
and by $X^*$ the set $X_1 \cap X_2$.
Then, if $S_1$ and $S_2$ hold, the D-sequent $S$ equal to \Dss{F}{X^*}{q}{X'} holds  too.
\end{proposition}
%
%
\begin{definition}
\label{def:res_rule}
We will say that the D-sequent $S$ of Proposition~\ref{prop:res_rule} is produced by \tb{resolving D-sequents}
 \pnt{S_1} \tb{and} \pnt{S_2}
 on variable $x$. $S$ is called the resolvent of $S_1$ and $S_2$ on  $x$.
\end{definition}
%
%
\subsection{Derivation of D-sequents in \Di}
\label{subsec:d_seq_dds_impl}
In this subsection, we discuss generation of D-sequents in \Di
(see Section~\ref{sec:dds_impl}). \Di builds a search tree by branching on
 variables of $X$ of $F(X,Y)$.
%
%
\begin{definition}
Let \pnt{q_1} and \pnt{q_2} be partial assignments to variables of $X$. We will denote by 
$\pnt{q_1} \leq \pnt{q_2}$ the fact that a) $\Va{q_1} \subseteq \Va{q_2}$ and b) every variable of \Va{q_1}
 is assigned in \pnt{q_1} exactly as in \pnt{q_2}.
\end{definition}

Let \pnt{q} be  the current  partial assignment to variables of $X$
and \Su{X}{red} be the  unassigned variables proved redundant in \cof{F}{q}. \Di generates a new D-sequent 
a) by resolving two existing D-sequents or b) if one of the conditions  below is true.

1) \ti{A (locally) empty clause appears in} \DDis{F}{q}{\Su{X}{red}}. Suppose, for example, that $F$ contains 
clause $C = x_1 \vee \overline{x_5} \vee x_7$.
Assume that assignments $(x_1\!=\!0,x_5\!=\!1)$ are made turning $C$ into the unit clause $x_7$.
 Assignment $x_7\!=\!0$ makes
$C$ an empty clause and so eliminates  all boundary points of  \DDis{F}{q}{\Su{X}{red}}. So \Di builds D-sequent
 \Dss{F}{\emptyset}{g}{X'} where \pnt{g}~=~($x_1\!=\!0,x_5\!=\!1,x_7\!=\!0$) and $X'$ is the set
 of unassigned variables
of \DDis{F}{q}{\Su{X}{red}} that are not in \Su{X}{red}.

2) \DDis{F}{q}{\Su{X}{red}} \ti{has only one  variable $x$} of $X$ that is not assigned and is not redundant.
In this case, \Di makes $x$ redundant by adding resolvents on variable $x$ and then builds D-sequent
 \DDs{F}{\Su{X'}{red}}{g}{x}
where $\Su{X'}{red} \subseteq \Su{X}{red}$, $\pnt{g} \leq \pnt{q}$ and   \Su{X'}{red} and \pnt{g}
 are defined in  Proposition~\ref{prop:d_seq_derivation} below
 (see also Remark~\ref{rem:general_idea}).

3) \ti{A monotone variable $x$ appears in formula} \DDis{F}{q}{\Su{X}{red}}. Then \Di builds 
D-sequent \DDs{F}{\Su{X'}{red}}{g}{x}
where $\Su{X'}{red} \subseteq \Su{X}{red}$, $\pnt{g} \leq \pnt{q}$ and   \Su{X'}{red} and \pnt{g} are defined in
  Proposition~\ref{prop:d_seq_derivation}
 (see Remark~\ref{rem:mon_var}).

Proposition~\ref{prop:d_seq_derivation} and Remark~\ref{rem:general_idea}
 below explain how to pick a subset of assignments of the
 current partial assignment \pnt{q} responsible
for the fact that a variable $x$  is redundant in branch \pnt{q}. This is similar to picking a subset of assignments
responsible for a conflict in SAT-solving.
%
%
\begin{proposition}
\label{prop:d_seq_derivation}
Let $F(X,Y)$ be a CNF formula and \pnt{q} be a partial assignment to variables of $X$. 
Let \Su{X}{red} be the variables proved  redundant in  \cof{F}{q}.
Let $x$ be the only variable of $X$ that is not in $\Va{q} \cup \Su{X}{red}$.
Let D-sequent \DDs{F}{\Su{X}{red}}{q}{x} hold.
Then D-sequent \DDs{F}{\Su{X'}{red}}{g}{x} holds where   \pnt{g} and $\Su{X'}{red}$ are defined as follows.
Partial assignment \pnt{g} to variables of $X$ satisfies   
the two  conditions below (implying that $\pnt{g} \leq \pnt{q}$):
\begin{enumerate}
\item Let $C$ be a \s{x}-clause of $F$ that is not in \DDis{F}{q}{\Su{X}{red}}. Then either
\label{enum:x_clauses}
  \begin{itemize}
   \item \pnt{g} contains an assignment satisfying $C$ or 
    \item  D-sequent  \DDs{F}{\Su{X^*}{red}}{g^*}{x^*} holds where
      $\pnt{g^*}~\leq~\pnt{g}$, $\Su{X^*}{red} \subset \Su{X}{red}$, $x^* \in (\Su{X}{red} \cap \V{C})$.
  \end{itemize}
\item Let \pnt{p_1} be a point such that  $\pnt{q} \leq \pnt{p_1}$. Let \pnt{p_1} 
\label{enum:non_x_clauses} 
falsify a clause of $F$ with literal $x$. Let \pnt{p_2} be obtained from \pnt{p_1} by 
flipping the value of $x$
and falsify a clause of $F$ with literal $\overline{x}$. Then there is a non-\s{x}-clause $C$ of $F$ falsified
by \pnt{p_1} and \pnt{p_2} such that  $(\V{C} \cap X) \subseteq \Va{g}$.
\end{enumerate}
The set \Su{X'}{red} consists of all the variables already proved redundant in \cof{F}{g}. That is every  redundant
variable $x^*$ of \Su{X}{red} with D-sequent \DDs{F}{\Su{X^*}{red}}{g^*}{x^*} such 
that $\pnt{g^*}~\leq~\pnt{g}$, $\Su{X^*}{red} \subset \Su{X}{red}$
is in \Su{X'}{red}.
\end{proposition}
%
\begin{remark}
\label{rem:general_idea}
When backtracking (and making new assignments)  formula \DDis{F}{q}{\Su{X}{red}} changes. 
Partial assignment \pnt{g} is formed so as to 
prevent  the changes that may produce  new \s{x}-boundary points. According
to Remark~\ref{rem:add_rem_clauses}, this may occur only in two cases. 

The first case is adding an \s{x}-clause $C$ to 
\DDis{F}{q}{\Su{X}{red}}. This may happen after backtracking if $C$ was satisfied or contained
 a redundant variable.
Condition~\ref{enum:x_clauses} of Proposition~\ref{prop:d_seq_derivation} makes \pnt{g} contain assignments 
 that prevent $C$ from appearing.

The second case is removing a non-\s{x}-clause $C$ from \DDis{F}{q}{\Su{X}{red}}. This may happen if 
 $C$ contains a literal falsified by an assignment in \pnt{q} and then this assignment is flipped.
 Condition~\ref{enum:non_x_clauses} of Proposition~\ref{prop:d_seq_derivation} 
makes \pnt{g} contain assignments guaranteeing that a ``mandatory'' set of clauses preventing 
 appearance of new \s{x}-boundary points
is present when D-sequent \DDs{F}{\Su{X'}{red}}{g}{x} is used.
\end{remark}
%
\begin{remark}
\label{rem:mon_var}
If  $x$ is monotone, Condition~\ref{enum:non_x_clauses} of Proposition~\ref{prop:d_seq_derivation} is vacuously true
because  \pnt{p_1} or \pnt{p_2} does not exist. So one can drop the requirement of 
Proposition~\ref{prop:d_seq_derivation} about  $x$ being the only variable of  $X$ that 
is not in $\Va{q} \cup \Su{X}{red}$. (It is used only when proving that the contribution of non-\s{x}-clauses into \pnt{g}
specified by Condition~\ref{enum:non_x_clauses} is correct. But if $x$ is monotone non-\s{x}-clauses are not used 
when forming \pnt{g}.)
\end{remark}
%
%
\subsection{Notation Simplification for D-sequents of \Di} 
\label{subsec:simp_notation}
In the description of \Di we will use the notation $\pnt{g} \rightarrow X''$
instead of $(F,X',\pnt{g}) \rightarrow X''$.  We do this for two reasons.
First, according to Proposition~\ref{prop:form_replacement}, in any D-sequent 
$(F_\mi{earlier},X',\pnt{g}) \rightarrow X''$, one can replace $F_\mi{earlier}$ with
$F_\mi{current}$ where the latter is obtained from the former by adding some resolvent-clauses.
Second, whenever \Di derives a new D-sequent, $X'$ is the set \Su{X}{red} of all unassigned variables of \cof{F}{q}
already proved redundant. So when we say that  $\pnt{g} \rightarrow X''$ holds we mean 
 that $(F,X',\pnt{g}) \rightarrow X''$
does  where $F$ is the current formula (i.e. the latest version of $F$) and $X'$ is $\Su{X}{red}$.

\section{Description of \Di}
\label{sec:dds_impl}
\subsection{Search tree}
  \Di branches on  variables of $X$ of $F(X,Y)$ building a search tree. 
  The current path of the search  tree is specified by partial assignment \pnt{q}. 
  \Di does not branch on variables proved  redundant for  current \pnt{q}.
 Backtracking to the root of the search tree means
derivation of D-sequent $\emptyset \rightarrow X$ (here we use the simplified notation of D-sequents, see
 Subsection~\ref{subsec:simp_notation}). At this point,
\Di terminates. 
We will denote the last variable assigned 
in \pnt{q} as \la{q}.

   Let $x$ be a branching variable. \Di maintains the notion of left and right branches corresponding
    to the first and second assignment to $x$ respectively. (In the modern SAT-solvers, the second
 assignment to a branching variable $x$
 is implied by  a clause $C$ derived in the left branch of $x$ where $C$ is empty in the left branch.
 A QEP-solver usually deals with satisfiable formulas.
If the left branch of $x$ contains a satisfying assignment, clause $C$ above does not exist.)

Although \Di distinguishes between decision and implied assignments (and  employs BCP procedure), 
no notion of decision levels
is  used. When an assignment (decision or implied) is made to a   variable, the depth of 
the current path increases by one
and  a new node of the search tree is created at the new depth. 
The current version of  \Di maintains a single search tree (no restarts are used). 
%
%
%
\subsection{Leaf Condition, Active D-sequents, Branch Flipping} 
\label{subsec:leaf_condition}
Every assignment made by \Di is added to \pnt{q}. The formula \Di operates on is  \DDis{F}{q}{\Su{X}{red}}. 
When a  monotone 
variable $x$ appears in \DDis{F}{q}{\Su{X}{red}}, it is added to the set 
 \Su{X}{red} of redundant variables of \cof{F}{q} and the \s{x}-clauses
are removed from \DDis{F}{q}{\Su{X}{red}}.
For every variable $x'$ of $\Su{X}{red}$ there is one  D-sequent \Ds{g}{x'} where $\pnt{g} \le \pnt{q}$.
We will call such a D-sequent \textbf{active}.
(Partial assignment \pnt{g} is in general different for different variables of $\Su{X}{red}$.) 
Let {\boldmath $\mi{D_\mi{seq}^{act}}$} denote the current set of active D-sequents. 

\Di keeps adding assignments to \pnt{q} until  every variable of $F$ is either assigned (i.e. in \Va{q}) or  
redundant (i.e. in \Su{X}{red}). We will refer to  this situation as \textbf{the leaf condition}.
The appearance of an empty clause in \DDis{F}{q}{\Su{X}{red}} is one of the cases where the leaf condition holds.

If  \Di is in the left branch of $x$ (where $x =\la{q}$) when the leaf condition occurs,
\Di starts the right branch by flipping the value of $x$. For every variable $x'$ of \Su{X}{red},
 \Di checks if \pnt{g} of D-sequent \Ds{g}{x'} contains an assignment to $x$. If it does, then
this D-sequent is not true any more. Variable $x'$ is removed from \Su{X}{red} and   \Ds{g}{x'}
is removed from \ADS and added to the set {\boldmath $\mi{D_\mi{seq}^{inact}}$} of inactive D-sequents. 
Every  \s{x'}-clause $C$ 
discarded from   \DDis{F}{q}{\Su{X}{red}} due to redundancy of $x'$ is recovered (unless $C$ contains
a variable that is still in \Su{X}{red}).

%
%
\subsection{Merging Results of Left and Right Branches}
\label{subsec:merging_branches}
If \Di is in the right branch of $x$ (where $x =\la{q}$) when the leaf condition occurs, then \Di does the following.
First \Di unassigns $x$.  Then \Di examines the list of variables removed from  \Su{X}{red}
after flipping the value of $x$. Let $x'$ be such a variable and $\Su{S}{left}$ and $\Su{S}{right}$ be the D-sequents
of $x'$ that were active in the left and right branch respectively. (Currently $\Su{S}{left}$ is in \IDS).
If $\Su{S}{right}$ does not depend on $x$, then $\Su{S}{left}$ is just removed from \IDS and $\Su{S}{right}$
remains in the set of active D-sequents \ADS. Otherwise, $\Su{S}{left}$ is resolved with $\Su{S}{right}$ on $x$.
Then $\Su{S}{left}$ and $\Su{S}{right}$ are removed from \IDS and \ADS respectively, and the resolvent is
added to \ADS and becomes a new active D-sequent of $x'$.

Then \Di makes variable $x$ itself redundant. (At this point every variable of $X$ but $x$ 
is either assigned or redundant.) To this end, \Di eliminates all \s{x}-removable
boundary points from \Fcurr by adding some resolvents on variable $x$. This is done as follows.
First, a CNF $H$ is formed from \Fcurr by removing all the \s{x}-clauses and adding a set 
of ``directing'' clauses \Su{H}{dir}. 
The latter is satisfied by an assignment \pnt{p} 
iff at least one clause $C'$ of \Fcurr with literal $x$ and
 one clause $C''$ with literal $\overline{x}$ is falsified by \pnt{p}. (How \Su{H}{dir} is built is
 described in~\cite{HVC}.) 
The satisfiability 
of $H$ is checked by calling a SAT-solver.
If  $H$ is satisfied by an assignment \pnt{p}, then the latter is an \s{x}-removable boundary point of
 \DDis{F}{q}{\Su{X}{red}}.
 It is eliminated 
by adding a resolvent $C$ on $x$ to \DDis{F}{q}{\Su{X}{red}}. (Clause $C$ is also added to $H$).
Otherwise, the SAT-solver returns a proof \ti{Proof} that $H$ is unsatisfiable.

Finally, a D-sequent \Ds{g}{x'} is generated satisfying the conditions of Proposition~\ref{prop:d_seq_derivation}.
To make \pnt{g} satisfy the second condition of Proposition~\ref{prop:d_seq_derivation}, \Di uses \ti{Proof}
above.  Namely,  every assignment falsifying a literal of a clause of \Fcurr used in \ti{Proof} 
is included in \pnt{g}.

%
%
%
\subsection{Pseudocode of \Di}
The main loop of \Di is shown in Figure~\ref{fig:dds_impl}. \Di can be in one of the six states
listed in Figure~\ref{fig:dds_impl}.  \Di terminates when it reaches the state \ti{Finish}.
Otherwise, \Di calls the procedure corresponding to the current state. This procedure performs
some actions and returns the next state of \Di\!\!. 

\Di starts in the \ti{BCP} state in which it runs the \ti{bcp} procedure (Figure~\ref{fig:bcp}).
Let $C$ be a unit clause of \DDis{F}{q}{\Su{X}{red}} where $\V{C} \subseteq X$.
As we mentioned in Subsection~\ref{subsec:d_seq_dds_impl},
\Di adds  D-sequent $\pnt{g} \rightarrow X''$ to \ADS where $X'' = X \setminus (\Va{q}~\cup~\Su{X}{red})$
and \pnt{g} is the minimal assignment falsifying $C$. This D-sequent corresponds to the (left) branch
of the search tree. In this branch, the only literal of $C$ is falsified, which makes the leaf condition true.

If a conflict occurs during BCP,  \Di switches to the state \ti{Conflict} and
calls a procedure that  generates a conflict clause \Su{C}{cnfl}
(Figure~\ref{fig:cnfl_processing}). Then \Di backtracks to the first node of the search tree at
 which \Su{C}{cnfl} becomes unit.

If BCP does not lead to a conflict, \Di switches to the state \ti{Decision\_Making} and calls a decision making procedure
(Figure~\ref{fig:dec_making}).
This procedure first looks for monotone variables. (\Su{X}{mon} of Figure~\ref{fig:dec_making}
 denotes the set of new monotone variables.)
If after processing monotone variables every unassigned
variable is redundant \Di switches to the \ti{Backtracking} state (and calls the \ti{backtrack}
 procedure, see Figure~\ref{fig:backtrack}).
Otherwise, a new assignment is made 
and added to \pnt{q}.

If \Di backtracks to the right branch of $x$ (where $x$ may be an implied or a decision variable),
 it switches to the state
BPE (Boundary Point Elimination) and calls the \ti{bpe} procedure (Figure~\ref{fig:bpe}). This procedure merges results
of left and right branches as described in Subsection~\ref{subsec:merging_branches}.
%
%
%
\subsection{Example}
\begin{example}
Let $F(X,Y)$ consist of  clauses: $C_1=x_1 \vee y_1$, $C_2=\overline{x}_1 \vee \overline{x}_2 \vee y_2$,
 $C_3=x_1 \vee x_2 \vee \overline{y}_3$. Let us consider how \Di builds formula $F^*(Y)$ equivalent 
to $\exists{X}.F(X,Y)$.
Originally, \pnt{q}, \Su{X}{red}, \ADS, \IDS are empty. Since $F$ does not have a unit clause, \Di
 switches to the  state \ti{Decision\_Making}.
Suppose \Di picks $x_1$ for branching and first
makes assignment $x_1=0$. At this point, $\pnt{q} = (x_1\!=\!0)$, clause $C_2$ is satisfied and
 \cof{F}{q} = $y_1 \wedge  (x_2 \vee \overline{y}_3)$.

Before making next decision, \Di processes the monotone variable $x_2$. First the  D-sequent \Ds{g}{x_2}
 is derived and added to \ADS
where $\pnt{g}=(x_1=0)$. (The appearance of the assignment $(x_1=0)$ in \pnt{g} is due to
Proposition~\ref{prop:d_seq_derivation}. According to Condition~\ref{enum:x_clauses}, 
\pnt{g} has to contain assignments that keep satisfied or redundant the \s{x_2}-
clauses that are not currently in \cof{F}{q}.
 The only \s{x_2}-clause that is not in \cof{F}{q} is $C_2$. It is satisfied by $(x_1=0)$.)
Variable $x_2$ is added to \Su{X}{red} and clause $x_2 \vee \overline{y}_3$ is removed from \cof{F}{q} 
as containing redundant variable $x_2$.
So \Fcurr = $y_1$.

Since $X$ has no variables to branch on (the leaf condition), \Di backtracks to the last assignment $x_1\!=\!0$
 and starts 
the right branch of $x_1$. So $\pnt{q}=(x_1\!=\!1)$. Since the D-sequent $(x_1=0) \rightarrow \s{x_2}$
 is not valid now, it is moved
from \ADS to $\mi{D_\mi{seq}^{inact}}$. Since $x_2$ is not redundant anymore it is removed from
 \Su{X}{red} and the clause $C_2$ is recovered
in \cof{F}{q} which is currently equal to $\overline{x}_2 \vee y_2$ (because $C_1$ and $C_3$ are satisfied by \pnt{q}).

Since $x_2$ is monotone again, D-sequent $(x_1=1) \rightarrow \s{x_2}$ is derived, $x_2$ 
is added to \Su{X}{red} and $C_2$ is
removed from \cof{F}{q}. So $\Fcurr = \emptyset$. At this point \Di backtracks to the right branch of 
$x_1$ and switches to the state \ti{BPE}.

In the \ti{BPE} state,  $x_1$ is unassigned.  $C_1$ satisfied by assignment $x_1=1$ is recovered. 
$C_2$ and $C_3$ (removed due to redundancy of $x_2$) are not recovered. The reason is that
redundancy of $x_2$ has been proved in both branches of $x_1$.  So $x_2$ stays redundant due to generation
 of  D-sequent
$\emptyset \rightarrow \s{x_2}$  obtained by  resolving D-sequents  $(x_1=0) \rightarrow \s{x_2}$
and $(x_1=1) \rightarrow \s{x_2}$
on $x_1$. So $\Fcurr = C_1$. D-sequent $\emptyset \rightarrow \s{x_2}$ replaces  
 $(x_1=1) \rightarrow \s{x_2}$ in  $\mi{D_\mi{seq}^{act}}$.
D-sequent  $(x_1=0) \rightarrow \s{x_2}$ is removed from  $\mi{D_\mi{seq}^{inact}}$. 

Then \Di is supposed to make $x_1$ redundant by adding resolvents on $x_1$ that eliminate
 \s{x_1}-removable boundary points of \DDis{F}{q}{\Su{X}{red}}. Since  $x_1$ is 
monotone in \DDis{F}{q}{\Su{X}{red}} it is already redundant.
So D-sequent $\emptyset~\rightarrow~\s{x_1}$ is derived and $x_1$ 
is added to \Su{X}{red}. Since
\pnt{q} is currently empty, \Di terminates returning an empty set of clauses as a  CNF formula  $F^*(Y)$ equivalent
to  $\exists{X}.F(X,Y)$.
\end{example}
\begin{proposition}
\label{prop:correctness}
\Di is sound and complete.
\end{proposition}
%
%
\setlength{\textfloatsep}{5pt}
\begin{figure}
\normalsize
// Given $F(X,Y)$, \Di  returns $F^*(Y)$ \\
 // such that $F^*(Y) \equiv \exists{X}.F(X,Y)$ \\
 // \pnt{q} is a partial assignment to vars of $X$ \\
// States of \Di are \ti{Finish}, \ti{BCP}, \ti{BPE},\\
// \ti{Decision\_Making}, \ti{Conflict}, \ti{Backtracking} \\ 
\vspace{-10pt}
\begin{tabbing}
a \=bb\= c\= \kill
\Di($F,X,Y$)\\
\>\{while (\ti{True}) \\
\Tt if (\ti{state} == \ti{Finish}) \\
\ttt return(\Dis{F}{X}); \\
\Tt if (\ti{state} == \ti{Non\_Finish\_State}) \\
\ttt \{\ti{state} = \ti{state\_procedure}(\pnt{q},\ti{other\_params}); \\
\ttt   continue;\}\} \\
\end{tabbing} 
\vspace{-20pt}
\caption{\textit{Main loop of \Di}}
\label{fig:dds_impl}
\end{figure}

%
%
\setlength{\floatsep}{10pt}
\begin{figure}
\normalsize
\begin{tabbing}
a \=bb\= c\= \kill
\textit{decision\_making}($\pnt{q},F,X,\Xr,\ADS$)\\
 \>\{($\Xm,\ADS(\Xm)) \leftarrow \mi{find\_monot\_vars}(F,X)$; \\
 \> $\ADS = \ADS(\Xr) \cup \ADS(\Xm)$; \\
 \> $\Xr = \Xr \cup \Xm$; \\
 \> if ($X == \Xr \cup \Va{q}$)  \\
 \Tt  if ($\Va{q} == \emptyset$) return(\ti{Finish}); \\
 \Tt else return(\ti{Backtracking}); \\
 \> $F$ = \Dis{F}{X_\mathit{mon}}; \\
 \> $\mi{assgn}(x) \leftarrow pick\_assgn(F,X)$;\\   
\> $\pnt{q'} = \pnt{q} \cup \mi{assgn}(x)$;\\
\> return(\ti{BCP});\} \\
\end{tabbing} 
\vspace{-20pt}
\caption{\textit{Pseudocode of the decision\_making procedure}}
\label{fig:dec_making}
\end{figure}

%
%
\begin{figure}
\normalsize
\begin{tabbing}
a \=bb\= c\= \kill
\ti{bcp}($\pnt{q},F,C_\mi{unsat}$)\\
\>\{$(\mi{answer},F,\pnt{q},C_\mi{unsat},\ADS)\leftarrow \mi{run\_bcp}(\pnt{q},F)$;\\
\> if ($\mi{answer}$ == \ti{unsat\_clause}) return(\ti{Conflict}); \\
\> else return(\ti{Decision\_Making});\}\\
\end{tabbing} 
\vspace{-20pt}
\caption{\textit{Pseudocode of the bcp procedure}}
\label{fig:bcp}
\end{figure}

%
%
\setlength{\floatsep}{10pt}
\begin{figure}
\normalsize
\begin{tabbing}
a \=bb\= c\= \kill
\ti{bpe}($\pnt{q},F,\Xr,\ADS,\IDS$)\\
\>\{$x = \la{q}$; \\
\> $(\pnt{q},F) \leftarrow unassign(\pnt{q},F,x)$; \\
\> $(F,\mi{Proof}) \leftarrow elim\_bnd\_pnts(F,x)$;\\
\> $\mi{optimize}(\mi{Proof})$; \\
\> $(\ADS,\IDS) \leftarrow \mi{resolve}(\ADS,\IDS,x)$;\\
\> $\ADS(\s{x}) = \mi{gen\_Dsequent}(\pnt{q},\mi{Proof})$; \\
\> $\ADS = \ADS(\Xr) \cup \ADS(\s{x})$; \\
\> $\Xr = \Xr \cup \s{x}$; \\
\> $F$ = \Dis{F}{\s{x}}; \\
\> if ($\Va{q} == \emptyset$) return(\ti{Finish});\\
\> else return(\ti{Backtracking});\} \\
\end{tabbing} 
\vspace{-20pt}
\caption{\textit{Pseudocode of the bpe procedure}}
\label{fig:bpe}
\end{figure}

%
%
\setlength{\textfloatsep}{7pt}
\begin{figure}
\normalsize
\begin{tabbing}
a \=bb\= c\= \kill
\ti{cnfl\_processing}($\pnt{q},F,C_\mi{unsat}$)\\
\>\{$(\pnt{q},F,C_\mi{cnfl}) \leftarrow \mi{gen\_cnfl\_clause}(\pnt{q},F,C_\mi{unsat})$; \\
\> $F = F \cup C_\mi{cnfl}$; \\
\> if ($C_\mi{cnfl} == \emptyset$) return(\ti{Finish}); \\
\> $x = \la{q}$; \\
\> if ($\mi{left\_branch}(x)$) return(\ti{BCP}); \\
\> else return(\ti{BPE});\}\\
\end{tabbing} 
\vspace{-20pt}
\caption{\textit{Pseudocode of the cnfl\_processing procedure}}
\label{fig:cnfl_processing}
\end{figure}

%
%
\setlength{\textfloatsep}{7pt}
\begin{figure}
\normalsize
\begin{tabbing}
a \=bb\= c\= \kill
\ti{backtrack}($\pnt{q},F,\Xr,\ADS,\IDS$)\\
\>\{$x = \la{q}$; \\
\> if ($\mi{right\_branch}(x)$) return(\ti{BPE});\\
\> $\pnt{q}= \mi{flip\_assignment}(\pnt{q},x)$; \\
\> $X' = \mi{find\_affected\_red\_vars}(\ADS(\Xr),x)$; \\
\> $\ADS = \ADS(\Xr) \setminus \ADS(X')$; \\
\> $\IDS = \IDS(\Xr) \cup \ADS(X')$; \\
\> $\Xr = \Xr \setminus X'$; \\
\> $F = \mi{recover\_clauses}(F,X')$; \\
\> return(\ti{BCP});\} \\
\end{tabbing}
\vspace{-20pt} 
\caption{\textit{Pseudocode of the backtrack procedure}}
\label{fig:backtrack}
\end{figure}

\section{Compositionality of \Di}
\label{sec:compos}
Let $F(X,Y)=F_1(X_1,Y_1) \wedge \ldots \wedge F_k(X_k,Y_k)$ where
$(X_i \cup Y_i) \cap (X_j \cup Y_j) = \emptyset$, $i \neq j$. As we mentioned in the introduction, 
the formula $F^*(Y)$ equivalent to $\exists{X}.F(X,Y)$ can be built 
as $F^*_1 \wedge \ldots \wedge F^*_k$ where 
$F^*_i(Y_i) \equiv \exists{X_i}.F_i(X_i,Y_i)$. 

We will say that a QEP-solver is \tb{compositional} if it reduces the problem of finding $F^*$ to $k$ independent
subproblems of building $F^*_i$. The DP-procedure~\cite{dp} is compositional (clauses of $F_i$ and $F_j$, $i \neq j$
cannot be resolved with each other). However, it may generate a huge number of redundant clauses.
A QEP-solver based on enumeration of satisfying assignments is not compositional. (The number of
blocking clauses, i.e. clauses eliminating satisfying assignments of $F$, is exponential in $k$). A QEP-solver
based on BDDs~\cite{bdds} is compositional but only for  variable orderings where
variables of $F_i$ and $F_j$, $i \neq j$ do not interleave. 
\begin{proposition}
\Di is compositional \ti{regardless} of  how branching variables are chosen. 
\end{proposition}

The fact that \Di is compositional regardless of branching choices is important in
practice. Suppose  $F(X,Y)$  does not have independent subformulas
but such subformulas appear in branches of the search tree.  A BDD-based QEP-solver may not be able to handle this case
because a BDD maintains one \ti{global} variable order (and different branches may require different variable orders).
\Di does not have such a limitation. It will \ti{automatically} use its compositionality
 whenever independent subformulas appear.

\section{Experimental Results}
\label{sec:experiments}
%
%
\begin{table}[htb]
\caption{Results for the sum-of-counters experiment}
\begin{center}
\begin{tabular}{|l|l|l|l|l|l|l|l|} \hline
 \#bits & \#coun-& \#state & \ti{Enum}- & Inter- & Inter- & \ti{DDS\_} &\ti{DDS\_} \\ 
        &  ters  & vars    &  \ti{SA}  & pol.   &  pol.  & \ti{impl}  &\ti{impl}  \\ 
        &        &         &  (s.) & \ti{Pico}.  &  \ti{Mini}. & rand.  & (s.) \\ 
        &        &         &       &  (s.)  &  (s.)  &  (s.) &      \\ \hline
    3   &   5    &    15   & 12.1   &   0.0  &  0.0   &  0.0  & 0.0  \\ \hline
    4   &  20    &    80   &$\ast$ &   0.4  &  0.1   &  0.5  & 0.4  \\ \hline
    5   &  40    &   200   &$\ast$ &   42   &  26    &   7   & 5  \\ \hline
    6   &  80    &   480   &$\ast$ & $\ast$ & $\ast$ &  101  & 67 \\ \hline
\end{tabular}
\label{tbl:sum_of_counters}
\end{center}
Instances marked with '$\ast$' exceeded the time limit (2 hours).
\end{table}
%

\setlength{\intextsep}{2pt}
\setlength{\floatsep}{2pt}
\begin{table}[htb]
\caption{Experiments with model checking formulas}
\begin{center}
\begin{tabular}{|p{32pt}|p{30pt}|p{10pt}|p{30pt}|p{12pt}|p{30pt}|p{12pt}|} \hline
  \Ss{model che-}  &  \multicolumn{2}{|p{40pt}|}{\ti{DP}} & \multicolumn{2}{|p{42pt}|}{\ti{EnumSA}} & \multicolumn{2}{|p{42pt}|}{\ti{DDS\_impl}} \\ 
\cline{2-3}\cline{4-5}\cline{6-7}
  \Ss{king mode}&  \Ss{solved}   &\Ss{time}      &\Ss{solved}      & \Ss{time}     &\Ss{solved}     & \Ss{time}  \\ 
        &\Ss{(\%)}  &   \Ss{(s.)}   & \Ss{(\%)}  &    \Ss{(s.)}  &\Ss{(\%)}  & \Ss{(s.)}   \\ \hline
\Ss{forward}   & \Ss{416 (54\%)}  &   \Ss{664}  &   \Ss{425 (56\%)}  &  \Ss{466}   & \Ss{531 (70\%)}  & \Ss{3,143}  \\ \hline
\Ss{backward}  & \Ss{47 (6\%)}   &   \Ss{13}   &   \Ss{97 (12\%)}   &  \Ss{143}   & \Ss{559 (73\%)}  & \Ss{690} \\ \hline

\end{tabular}
\label{tbl:model_checking}
\end{center}
The time limit is 1 min.
\end{table}
In this section, we give results of some experiments with an implementation of \ti{DDS\_impl}. 
The objectives of our experiments were a) to emphasize the compositionality of \ti{DDS\_impl}; b)
 to compare \ti{DDS\_impl} with a QEP-solver based on enumeration of
satisfying assignments. As a such QEP-solver we used an implementation of the algorithm recently
introduced at CAV-11~\cite{cav11} (courtesy of Andy King).  (We will refer to this QEP-solver as \ti{EnumSA}). For the sake
of completeness we also compared \Di and \ti{EnumSA} with our implementation of the DP procedure.

Our current implementation of \Di
is not particularly well optimized yet and written just to satisfy the two objectives above.  For example,
to simplify the code, the SAT-solver employed to find boundary points does not use fast BCP
(watched literals). More importantly, the current version of \Di lacks important features that should 
have a dramatic impact on its performance. For example, to simplify memory management, \Di
 does not currently  \ti{reuse} D-sequents. As soon
as two D-sequents are resolved (to produce a new D-sequent) they are discarded.

To verify the correctness of results of \Di we used two approaches. If an instance $\exists{X}.F(X,Y)$
 was solved by \ti{EnumSA}
we simply checked  the CNF formulas $F^*(Y)$ produced by \Di and \ti{EnumSA} for equivalence. Otherwise,
we applied a two-step procedure. First, we checked that every clause of $F^*$ was implied by $F$.
Second, we did random testing to see if $F^*$ missed some clauses. Namely, we randomly generated
assignments \pnt{y} satisfying $F^*$. For every \pnt{y} we checked if it could be extended
to (\pnt{x},\pnt{y}) satisfying $F$. (If no such extension exists, then $F^*$ is incorrect.)

In the first experiment (Table~\ref{tbl:sum_of_counters}), we considered a circuit $N$ 
of $k$ independent $m$-bit counters. 
Each counter
had an independent input variable. The property  we checked (further referred to as $\xi$) was
$\mi{Num}(\mi{Cnt}_1) + \ldots + \mi{Num}(\mi{Cnt}_k) < R$. 
 Here $\mi{Num}(\mi{Cnt}_i)$ is the number specified
by the outputs of $i$-th counter and  $R$ is a constant equal to $k \ast (2^m-1)+1$. Since, the maximum number
that appears at the outputs of a counter is $2^m - 1$, property $\xi$ holds. Since the counters are independent
of each other, the state space of $N$ is the Cartesian product of the $k$ state spaces of individual counters.
However, property $\xi$ itself is not compositional (one cannot verify it by solving $k$-independent
 subproblems), which makes verification harder.

The first two columns of Table~\ref{tbl:sum_of_counters} give the value of $m$ and $k$ of four circuits $N$.
The third column specifies the number of state variables (equal to $m \ast k$). In this experiment, we applied \ti{EnumSA}
and \Di to verify property $\xi$ using forward model checking. In either case, the QEP-solver was used to
compute  CNF formula $\mi{RS^*}(\Su{S}{next})$ specifying the next set of reachable states. It was obtained
from  formula
 $\exists{\Su{S}{curr}}\exists{X}.\mi{Tr}(\Su{S}{curr},\Su{S}{next},X) \wedge \mi{RS}_p(\Su{S}{curr})$ by quantifier 
elimination.
Here \ti{Tr} is a CNF formula representing the transition relation and $\mi{RS}_p(\Su{S}{curr})$ specifies the set of
 states reached in $p$ iterations.
$\mi{RS}_{p+1}(\Su{S}{curr})$ was computed as a CNF formula equivalent to
 $\mi{RS}_p(\Su{S}{curr}) \vee \mi{RS^*}(\Su{S}{curr})$.

We also estimated the complexity of verifying  the examples of Table~\ref{tbl:sum_of_counters} by
interpolation~\cite{interpolation}. Namely, we used \ti{Picosat~913} and \ti{Minisat~2.0} for finding a 
proof that $\xi$ holds for $2^{m-1}$ iterations
(the diameter of circuits $N$ of Table~\ref{tbl:sum_of_counters} is $2^m$, $m=3,4,5,6$). Such a proof is used in the
  method of~\cite{interpolation}
to extract an interpolant. So, in Table~\ref{tbl:sum_of_counters}, we give only the time necessary to find the first 
interpolant.

Table~\ref{tbl:sum_of_counters} shows that \ti{EnumSA} does not scale well (the number of blocking clauses one has 
to generate for the formulas of 
Table~\ref{tbl:sum_of_counters}  is exponential in the number of counters). Computation of interpolants scales much
 better,
 but  \ti{Picosat} and \ti{Minisat} failed to compute a proof for the largest example in 2 hours.

The last two columns of Table~\ref{tbl:sum_of_counters} give  the performance of \Di when branching variables were
 chosen randomly (next to last column)
and heuristically (last column). In either case, \Di shows good scalability  explained by  the fact that  \Di is 
compositional. Moreover,
the fact that the choice of branching variables is not particularly important means that \Di has a ``stronger''
 compositionality 
than  BDD-based QEP-solvers. The latter are compositional only for particular variable orderings.

In second and third experiments (Table~\ref{tbl:model_checking}) we used the 758 model checking benchmarks of HWMCC'10 competition~\cite{hwmcc10}. In the second
experiment, (the first  line of Table~\ref{tbl:model_checking}) we used \ti{DP}, \ti{EnumSA} and \Di to compute 
the set of states reachable in the first transition. In
this case  one needs to find CNF formula $F^*(Y)$ equivalent to $\exists{X}.F(X,Y)$ where 
$F(X,Y)$ specifies the transition relation  and the initial state.  Then $F^*(Y)$ 
gives  the set of states reachable in one transition.

In the third experiment, (the second line of Table~\ref{tbl:model_checking}) we used the same benchmarks to compute the set of bad states
in backward model checking. In this case, formula $F(X,Y)$ specifies the output function and the property (where $Y$ is the set of variables describing
the current state). If $F(X,Y)$ evaluates to 1 for some assignment (\pnt{x},\pnt{y}) to $X \cup Y$,  the property is broken and the state
specified by \pnt{y} is ``bad''. The formula $F^*(Y)$ equivalent to $\exists{X}.F(X,Y)$ specifies  the set of bad states.

Table ~\ref{tbl:model_checking} shows the number  of benchmarks solved by each program and the percentage of this number  to 758.
Besides the time taken by each program for the \ti{solved} benchmarks is shown.  \Di solved more benchmarks than \ti{EnumSA}
and \ti{DP} in forward model checking and dramatically more benchmarks in the backward model checking. \Di needed more time
than  \ti{DP} and \ti{EnumSA} because typically the benchmarks  solved only by \Di were the most time consuming. 

\section{Background}
\label{sec:background}
The notion of boundary points was  introduced in~\cite{date02}. 
for pruning the search tree (in the context
of SAT-solving).  The relation between a resolution proof and  the process of elimination
of boundary points was discussed in~\cite{bnd_pnts,HVC}. The previous papers considered 
 only the notion of \s{z}-boundary
of formula $G(Z)$ where $z$ is a variable of $Z$. In the present paper, we consider $Z'$-boundary points
where $Z'$ is  an arbitrary subset of $Z$. (This extension is not trivial and at the same time
crucial for the introduction of D-sequents.)

The idea of a QEP-solver based on enumerating satisfying assignments was introduced in~\cite{blocking_clause}. 
It has been further developed in ~\cite{fabio,cofactoring,cav11}.  In~\cite{interpolation} it was shown how one can avoid
QEP-solving in reachability analysis by building interpolants. Although, this direction is very promising,
interpolation based methods have to overcome the following problem. In the current implementations, interpolants are extracted
 from resolution proofs.
 Unfortunately, modern SAT-solvers
are still not good enough to take into account the high-level structure of a formula. (An example of that
is given in Section~\ref{sec:experiments}.) So proofs they find and the interpolants extracted from those
proofs may have poor quality.

Note that our notion of  redundancy of variables is different from observability related notions of redundancy.
For instance, in contrast to the notion of careset~\cite{careset}, if a CNF formula $G(Z)$ is satisfiable,
\ti{all} the variables of $Z$ are redundant in the formula  $\exists{Z}.G(Z)$ according to our definition. 
($G$ may have a lot of boundary points, but none of them is removable.
So   $\exists{Z}.G(Z)$ is equivalent to an empty CNF formula. Of course, to prove the variables of  $Z$ redundant,
one has to derive  D-sequent $\emptyset \rightarrow Z$.)

\section{Conclusion}
\label{sec:conclusion}
We present a new method for eliminating existential quantifiers from a Boolean CNF formula $\exists{X}.F(X,Y)$.
The essence of this method is to add resolvent clauses to $F$ and record the decreasing dependency
on variables of $X$ by dependency sequents  (D-sequents).  An algorithm based on this method 
(called DDS, Derivation of D-Sequents) terminates when it derives the D-sequent saying that the variables
of $X$ are redundant. Using this termination condition  may lead to a significant performance improvement
in comparison to the algorithms based on enumerating satisfying assignments.
This improvement   may be even
exponential (e.g. if a CNF formula is composed of independent subformulas.)

 Our preliminary experiments with a very simple implementation show the  promise of DDS. 
At the same time, DDS  needs further study.  Here are some directions for future research:
a) decision making heuristics; b) reusing D-sequents; c) efficient data structures;
d) getting information about the structure of the formula (specified as  a sequence of D-sequents to derive).

\section{Acknowledgment}
This work was funded  in part by NSF grant CCF-1117184 and  SRC contract 2008-TJ-1852.

\appendix
\setcounter{proposition}{0}

%
%
\section*{Proofs of Section~\ref{sec:basic}}
%
%
\begin{proposition}
Point \pnt{p} is a $Z'$-removable boundary point of a CNF formula $G(Z)$
iff no point \pnt{p^*} obtained from \pnt{p} by changing values of (some) variables of $Z'$ satisfies $G$.
\end{proposition}
\begin{mmproof}
\textit{If part.}
Let us partition $G$ into $G_1$ and $G_2$ where $G_1$ is the set of $Z'$-clauses and $G_2$ is the
set of of non-$Z'$-clauses. By definition, \pnt{p} is a $Z''$-boundary point where $Z'' \subseteq Z'$.
So \pnt{p} satisfies $G_2$. 

 Let $C$ be the clause such that 
\begin{itemize}
\item $\V{C} = Z \setminus Z'$,
\item $C$ is falsified by \pnt{p}.
\end{itemize}

Clause $C$ is implied by $G_1$. Indeed, assume the contrary i.e.  there exists  \pnt{p^*} for which $G_1$(\pnt{p^*})=1
 and $C$(\pnt{p^*})=0.  Note that since \pnt{p^*} falsifies $C$, it can be different from \pnt{p} only
 in assignments to $Z \setminus Z'$.
Then, there is a point \pnt{p^*} obtained by flipping values of $Z'$ 
that satisfies  $G_1$. But since \pnt{p^*} has the same assignments to variables of $Z \setminus Z'$
as \pnt{p}, it satisfies $G_2$ too. So \pnt{p^*} is obtained by flipping assignments of $Z'$ and satisfies $G$,
 which contradicts the assumption 
of the proposition at hand. So $C$ is implied by $G_1$.
Since $C$ satisfies the conditions of Definition~\ref{def:rem_bnd_pnt}, \pnt{p} is a $Z'$-removable
boundary point.

\vspace{6pt}
\noindent
\textit{Only if part.} Assume the contrary. That is there is clause $C$ satisfying the conditions
of Definition~\ref{def:rem_bnd_pnt} and there is a point \pnt{p^*} obtained from \pnt{p} by flipping
values of variables of $Z'$ that satisfies $G$. Then \pnt{p^*} also satisfies the set $G_1$ of
$Z'$-clauses of $G$.  Since $C$ is implied by $G_1$, then $C$ is satisfied by \pnt{p^*} too.
Since \pnt{p} and \pnt{p^*} have identical assignments to the variables of $Z \setminus Z'$, then
$C$ is also satisfied by \pnt{p}. However this contradicts one of the conditions of Definition~\ref{def:rem_bnd_pnt}
assumed to be true.
\end{mmproof}
%
%
\section*{Proofs of Section~\ref{sec:red_vars}}
%
%
\begin{lemma}
\label{lemma:twin_bnd_pnts}
Let \pnt{p'} be a \s{z}-boundary point of CNF formula $G(Z)$ where $z \in Z$.
Let \pnt{p''} be obtained from \pnt{p'} by flipping the value of $z$.
Then \pnt{p''} either satisfies $F$ or it is also a \s{z}-boundary point.
\end{lemma}
\begin{mmproof}
Assume the contrary i.e. \pnt{p''} falsifies a clause $C$ of $G$ that does not
have a variable of $z$. (And so  \pnt{p''} is neither a satisfying assignment
nor a \s{z}-boundary point of $G$.) Since \pnt{p'} is different from \pnt{p''}
only in the value of $z$, it also falsifies $C$. Then \pnt{p'} is not
a \s{z}-boundary point of $G$. Contradiction.
\end{mmproof}
%
%
\begin{proposition}
  Let $G(Z)$ be a CNF formula and $z$ be a monotone variable of $F$. (That is clauses of $G$ contain the literal of $z$
 of only one polarity.)  Then  $z$ is redundant in $G$.
\end{proposition}
\begin{mmproof}
Let us consider the following two cases.
\begin{itemize}
\item $G(Z)$ does not have a \s{z}-boundary point. Then the proposition holds.
\item $G(Z)$ has a \s{z}-boundary point \pnt{p'}. Note that the clauses of $G$ falsified by \pnt{p'} have the same
literal $l(z)$ of variable $z$. Let \pnt{p''} be the point obtained from \pnt{p'} by flipping the value of $z$.
According to Lemma~\ref{lemma:twin_bnd_pnts}, one needs to consider only the following two cases.
   \begin{itemize}
     \item \pnt{p''} satisfies $G$. Then \pnt{p'} is not a \s{z}-removable boundary point. This implies that \pnt{p'} is not
      a removable boundary point of $G$ either (see Remark~\ref{rem:bnd_pnt_removability}). So the proposition holds.
     \item \pnt{p''} falsifies only the clauses of $G$ with literal $\overline{l(z)}$. (Point \pnt{p''}
       cannot falsify a clause with literal $l(z)$.) Then $G$ has literals of $z$ of both polarities and $z$
       is not a monotone variable. Contradiction.
   \end{itemize}
\end{itemize}
\end{mmproof}

%
%
\begin{proposition}
Let  $F(X,Y)$ be a CNF formula and $X'$ be a subset of  $X$.
Then $\exists{X}.F(X,Y) \equiv \exists{(X \setminus X')}.\Dis{F}{X'}$ iff the variables of $X'$ are redundant
in $F$.
\end{proposition}
\begin{mmproof}
Denote by $X''$ the set $X \setminus X'$ and by $F^*(X'',Y)$ the formula \Dis{F}{X'}.

\noindent\textit{If part.} 
Assume the contrary i.e. the variables of $X'$ are redundant but  
$\exists{X}.F(X,Y) \not\equiv \exists{X''}.F^*(X'',Y)$.  Let \pnt{y} be an assignment to $Y$ such that
$\exists{X}.F(X,\pnt{y}) \neq  \exists{X''}.F^*(X'',\pnt{y})$.
One has to consider the following two cases. 
\begin{itemize}
\item $\exists{X}.F(X,\pnt{y})=1$, $\exists{X''}.F^*(X'',\pnt{y})=0$.  
Then there exists 
an assignment \pnt{x} to $X$ such that (\pnt{x},\pnt{y}) satisfies $F$. Since every clause of $F^*$
is in $F$, formula $F^*$ is also satisfied by (\pnt{x''},\pnt{y}) where \pnt{x''} consists of the assignments
of \pnt{x} to variables of $X''$. Contradiction.
\item $\exists{X}.F(X,\pnt{y})=0$, $\exists{X''}.F^*(X'',\pnt{y})=1$. Then there exists an assignment \pnt{x''}
to variables of $X''$ such that (\pnt{x''},\pnt{y}) satisfies $F^*$. Let \pnt{x} be an assignment to $X$
obtained from \pnt{x''} by arbitrarily assigning variables of $X'$. Since $F(X,\pnt{y})\equiv 0$,
point (\pnt{x},\pnt{y}) falsifies $F$. Since $F^*(\pnt{x},\pnt{y})=1$ and every clause of $F$ that
is not $F^*$ is an $X'$-clause, (\pnt{x},\pnt{y}) is an $X'^*$-boundary point of $F$.  Since $F(X,\pnt{y})\equiv 0$,
(\pnt{x},\pnt{y}) is removable. Hence the variables of $X'$ are not redundant in $F$. Contradiction.
\end{itemize}
\vspace{6pt}
\noindent
\textit{Only if part.} Assume the contrary i.e. $\exists{X}.F(X,Y) \equiv \exists{X''}.F^*(X'',Y)$ but the variables
of $X'$ are not redundant in $F$. Then there is an $X'^*$ boundary point \pnt{p}=(\pnt{x},\pnt{y})
 of $F$ where $X'^* \subseteq X'$
that is removable in $F$. Since \pnt{p} is a boundary point, $F(\pnt{p}) = 0$. Since
\pnt{p} is removable, $\exists{X}.F(X,\pnt{y}) = 0$. On the other hand, since \pnt{p}
falsifies only $X'$-clauses of $F$, it satisfies $F^*$. Then the point (\pnt{x''},\pnt{y}) obtained from \pnt{p}
by dropping the assignments to $X'$ satisfies $F^*$. Hence $\exists{X''}.F^*(X'',\pnt{y}) = 1$ and so
$\exists{X}.F(X,\pnt{y}) \neq \exists{X''}.F^*(X'',\pnt{y})$. Contradiction.
\end{mmproof}

%
%
\section*{Proofs of Section~\ref{sec:add_remove_clauses}}
\begin{definition}
Point \pnt{p} is called a \pnt{Z'}\textbf{-unremovable boundary point} of $G(Z)$ where $Z' \subseteq Z$ if 
\pnt{p} is a $Z''$-boundary point where $Z'' \subseteq Z'$ and  clause $C$ of Definition~\ref{def:rem_bnd_pnt}
does not exist. (According to Proposition~\ref{prop:remv_bpnt} this means that by flipping values of variables 
of $Z'$ in \pnt{p}
one can get a point satisfying $G$.)
\end{definition}
%
%
\begin{definition}
Let $G(Z)$ be a CNF formula and  \pnt{p} be an $Z'$-boundary point of $G$ where $Z' \subseteq Z$.
A point \pnt{p^*} is called a $Z''$-neighbor of \pnt{p} if
\begin{itemize}
\item $Z' \subseteq Z''$
\item \pnt{p} and \pnt{p^*} are different only in (some) variables of $Z''$. In other words, \pnt{p} and \pnt{p^*}
 can be obtained from each other by flipping (some) variables of $Z''$. 
\end{itemize}
\end{definition}
\vspace{4pt}
%
%
\begin{proposition}
Let $G(Z)$ be a CNF formula. Let $G$ have no \s{z}-removable boundary points.
Let $C$ be a clause. Then the formula $G \wedge C$ does not have a \s{z}-removable boundary point 
if at least one of the following conditions hold: \linebreak a) $C$ is implied by $G$; b) $z \notin$ \V{C}.
\end{proposition}
\begin{mmproof}
Let \pnt{p} be a complete assignment to the variables of $G$ (a point) and $C$ be a clause
satisfying at least one of the two conditions of the proposition.
 Assume the contrary i.e. that
 \pnt{p}  is a \s{z}-removable boundary point of $G \wedge C$.

Let us consider the following four cases.
\begin{enumerate}
\item $G$(\pnt{p})=0, $C$(\pnt{p})=0. 
  \begin{itemize}
   \item Suppose that  \pnt{p} is not a \s{z}-boundary point of $G$. Then  it falsifies a clause $C'$
of $G$ that is not a \s{z}-clause. Then \pnt{p} is not a \s{z}-boundary point of $G \wedge C$. Contradiction.
   \item Suppose that \pnt{p} is a \s{z}-unremovable boundary point of $G$. (According to the conditions of 
the proposition at hand,
    $G$ cannot have a \s{z}-removable boundary point.) This means that the point
    \pnt{p'} that is the \s{z}-neighbor of \pnt{p} satisfies $G$.
    \begin{itemize}
     \item Assume that $C$ is not a \s{z}-clause. Then \pnt{p} is not a 
            \s{z}-boundary point of $G \wedge C$. Contradiction.
     \item Assume that $C$ is implied by $G$. Then  $C$(\pnt{p'})=1 and so \pnt{p'} satisfies  $G \wedge C$.
              Then \pnt{p} is still a \s{z}-unremovable boundary point of $G \wedge C$. Contradiction.
     \end{itemize}
  \end{itemize}
\item $G$(\pnt{p})=0, $C$(\pnt{p})=1. 
  \begin{itemize}
   \item Suppose that  \pnt{p} is not a \s{z}-boundary point of $G$. Then  it falsifies a clause $C'$
of $G$ that is not a \s{z}-clause. Then \pnt{p} is not a \s{z}-boundary point of $G \wedge C$. Contradiction.
   \item Suppose that \pnt{p} is a \s{z}-unremovable boundary point of $G$. This means that the point
    \pnt{p'} that is the \s{z}-neighbor of \pnt{p} satisfies $G$.
    \begin{itemize}
     \item Assume that $C$ is not a \s{z}-clause. Then $C$(\pnt{p})=$C$(\pnt{p'}) 
                and so $C$(\pnt{p'})=1. Then \pnt{p'} satisfies
        $G \wedge C$ and so \pnt{p} is a \s{z}-unremovable boundary point of $G \wedge C$. Contradiction.
     \item Assume that $C$ is implied by $G$ and so $C$(\pnt{p'})=1. Hence \pnt{p'}
            satisfies $G \wedge C$.
              Then \pnt{p} is a \s{z}-unremovable boundary point of $G \wedge C$. Contradiction.
     \end{itemize}
  \end{itemize}
\item $G$(\pnt{p})=1, $C$(\pnt{p})=0. 
   \begin{itemize}
   \item If $C$ is implied by $G$, then we immediately get a contradiction.
   \item If $C$ is not a \s{z}-clause, then \pnt{p} falsifies a non-\s{z}-clause of $G \wedge C$   and
      so \pnt{p} is not a \s{z}-boundary point of $G \wedge C$. Contradiction.
   \end{itemize}
\item $G$(\pnt{p})=1, $C$(\pnt{p})=1. Point \pnt{p} satisfies $G \wedge C$ and so cannot be a \s{z}-boundary point
of $G \wedge C$. Contradiction.
\end{enumerate}
\end{mmproof}

%
%
\begin{proposition}
Let $G(Z)$ be a CNF formula. Let $G$ have no \s{z}-removable boundary points.
Let $C$ be a \s{z}-clause of $G$. Then the formula $G' = G \setminus \{C\}$ does not have a
\s{z}-removable boundary point.
\end{proposition}
\begin{mmproof}
Let \pnt{p} be a complete assignment to the variables of $G$ (a point).  Assume the contrary i.e. that
$z \in$ \V{C} and \pnt{p} is a \s{z}-removable boundary point of $G'$.
Let us consider the following three cases.
\begin{enumerate}
\item $G$(\pnt{p})=0, $C$(\pnt{p})=0. 
   \begin{itemize} 
   \item Suppose that \pnt{p} is not a \s{z}-boundary point of $G$. Then there is clause $C'$ of $G$ that is
not a \s{z}-clause  and that is falsified by \pnt{p}. Since $C'$ is different from $C$ (because the former
is not a \s{z}-clause) it remains in $G'$. Hence \pnt{p} is not a \s{z}-boundary point of $G'$. Contradiction.
   \item Suppose that \pnt{p} is a \s{z}-unremovable boundary point of $G$. Then its \s{z}-neighbor \pnt{p'} satisfies
    $G$ and hence $G'$. Then \pnt{p} either satisfies $G'$ (if $C$ is the only \s{z}-clause of $G$ falsified by \pnt{p})
   or \pnt{p} is a \s{z}-unremovable boundary point of $G'$. In either case, we have a contradiction.
   \end{itemize}
\item $G$(\pnt{p})=0, $C$(\pnt{p})=1.
    \begin{itemize}
    \item Suppose that \pnt{p} is not a \s{z}-boundary point of $G$. Using the same reasoning as above we get a
          contradiction.
    \item  Suppose that \pnt{p} is a \s{z}-unremovable boundary point of $G$. Then its \s{z}-neighbor \pnt{p'} satisfies
    $G$ and hence $G'$. Let $C'$ be a \s{z}-clause of $G$ falsified by \pnt{p}. Since $C'$ is different from $C$
   (the latter
    being satisfied by \pnt{p}), it is present in $G'$. Hence \pnt{p} falsifies $G'$.
    Then \pnt{p}  is a \s{z}-unremovable boundary point of $G'$. We have a contradiction.
    \end{itemize} 
\item $G$(\pnt{p})=1. Then $G'$(\pnt{p})=1 too and so \pnt{p} cannot be a boundary point of $G'$. Contradiction.
\end{enumerate}
\end{mmproof}
%
%
\section*{Proofs of Section~\ref{sec:d_sequents}}
\subsection*{SUBSECTION: Formula Replacement in a D-sequent}
\begin{proposition}
Let $F^+(X,Y)$ be a CNF formula obtained from $F(X,Y)$ by adding some resolvents of clauses of $F$.
Let \pnt{q} be a partial assignment to variables of $X$ and $X' \subseteq X$.
Then the fact that D-sequent $(F,X',\pnt{q}) \rightarrow X''$ holds implies that
 $(F^+,X',\pnt{q}) \rightarrow X''$ holds too. The opposite is not true.
\end{proposition}
\begin{mmproof}
First, let us prove that if $(F,X',\pnt{q}) \rightarrow X''$ holds, $(F^+,X',\pnt{q}) \rightarrow X''$ holds too.
Let us assume the contrary, i.e. $(F,X',\pnt{q}) \rightarrow X''$ holds but $(F^+,X',\pnt{q}) \rightarrow X''$ does not.
According to Definition~\ref{def:d_sequent}, this means  that either 
\begin{itemize}
\item[A)] variables of $X'$ are not redundant in \cof{F^+}{q} or 
\item[B)] variables of $X''$ are not redundant in \DDis{F^+}{q}{X'}.
\end{itemize}

\noindent{CASE A:}
The fact that the variables of $X'$ are not redundant in \cof{F^+}{q}  means
that there is a removable $X'^*$-boundary point \pnt{p}  of \cof{F^+}{q} where $X'^* \subseteq X'$. The fact that the variables of $X'$ are
redundant in \cof{F}{q} means that \pnt{p} is not a removable $X'^*$-boundary point of \cof{F}{q}.
Let us consider the  three reasons  for that.
\begin{itemize}
\item \pnt{p} satisfies \cof{F}{q}. Then it also satisfies \cof{F^+}{q} and hence cannot be a boundary point of
 \cof{F^+}{q}. Contradiction.
\item \pnt{p} is not  an $X'^*$-boundary point of \cof{F}{q}. That is \pnt{p} falsifies a
non-$X'$-clause $C$ of \cof{F}{q}.
 Since \cof{F^+}{q} also contains $C$, 
point \pnt{p} cannot
              be an $X'^*$-boundary point of \cof{F^+}{q} either. Contradiction.
\item \pnt{p} is an $X'^*$-boundary point of \cof{F}{q} but  it is not removable. This means that one can 
obtain a point \pnt{p^*}
              satisfying \cof{F}{q} by flipping the values of variables of $X \setminus \Va{q}$ in \pnt{p}. Since \pnt{p^*} 
also satisfies \cof{F^+}{q},
              one has to conclude that \pnt{p} is not a removable point of \cof{F^+}{q}. Contradiction.
\end{itemize}

\noindent{CASE B:}
The fact that  the variables of $X''$ are not redundant in  \DDis{F^+}{q}{X'} means
that there is a removable $X''^*$-boundary point \pnt{p}  of \DDis{F^+}{q}{X'} where $X''^* \subseteq X''$. The fact that the variables of $X''$ are
redundant in \DDis{F}{q}{X'} means that \pnt{p} is not a removable $X''^*$-boundary point of \DDis{F}{q}{X'}.

Here one can reproduce the reasoning of case A). That is one can consider the three cases above describing
 why \pnt{p}  is not an removable $X''^*$-boundary point of  \DDis{F}{q}{X'} and show that each case leads to a 
contradiction for the same reason as above.

\vspace{10pt}
Now we show that if $(F^+,X',\pnt{q}) \rightarrow X''$ holds this does not mean that 
 $(F,X',\pnt{q}) \rightarrow X''$ holds too. 
Let $F(X,Y)$ be a CNF formula where $X=\s{x}, Y=\s{y}$. Let $F$ consist of clauses $C_1$,$C_2$
 where $C_1=x \vee y$ and $C_2 =\overline{x} \vee y$.
Let $F^+$ be obtained from $F$ by adding the unit clause $y$ (that is the resolvent of $C_1$ 
and $C_2$). It is not hard to see
that the D-sequent $(F^+,\emptyset,\emptyset) \rightarrow \s{x}$ holds. (The latter does not 
have any \s{x}-boundary points. Hence it cannot have
a removable \s{x}-boundary point.) At the same time, $F$ has a removable \s{x}-boundary point 
\pnt{p}=($x$=0,$y$=0). So the D-sequent
  $(F,\emptyset,\emptyset)~\rightarrow~\s{x}$ does not hold.
\end{mmproof}

\subsection*{SUBSECTION: Resolution of D-sequents}
%
%
\begin{definition}
\label{def:loc_red_vars}
Let $F(X,Y)$ be a CNF formula and $X'~\subseteq~X$. We will say that the variables of $X'$
are \tb{locally redundant} in $F$ if every $X''$-boundary point \pnt{p} of $F$ where $X'' \subseteq X'$
is $X'$-removable.
\end{definition}
%
%
\begin{remark}
\label{rem:difference}
We will call  the  variables of a set $X'$  \tb{globally redundant} in $F(X,Y)$ if they are redundant
in the sense of Definition~\ref{def:glob_red_vars}. The difference between locally and globally redundant
variables is as follows. When testing if variables of $X'$ are redundant, in either case one checks
if every $X''$-boundary point \pnt{p} of $F$ where $X'' \subseteq X'$ is removable. The difference is in the set
variables one is allowed to change. In the case of locally redundant variables (respectively globally redundant
variables) one checks if \pnt{p} is $X'$-removable (respectively $X$-removable). In other words, in the case of globally
variables one is allowed to change variables that are not in $X'$.
\end{remark}
\begin{lemma}
\label{lemma:special_case}
If variables of $X'$ are locally redundant in a CNF formula $F(X,Y)$ they are also globally redundant there.
The opposite is not true.
\end{lemma}
\begin{mmproof}
See Remark~\ref{rem:difference}.
\end{mmproof}
%
%
\begin{lemma}
\label{lemma:mon_loc_red}
Let $z$ be a monotone variable of $G(Z)$. Then variable $z$ is \ti{locally} redundant.
\end{lemma}
\begin{mmproof}
Let us assume for the sake of clarity that only positive literals of $z$ occur in clauses of $G$.
Let us consider the following two cases:
\begin{itemize}
\item Let $G$ have no any \s{z}-boundary points. Then the proposition is vacuously true.
\item  Let \pnt{p} be a \s{z}-boundary point. By flipping the value of $z$ from 0 to 1, we obtain
 an assignment satisfying $G$. So \pnt{p}  is not a removable \s{z}-boundary point and to prove
that it is sufficient to flip the value of $z$. Hence $z$ is locally redundant in $G$.
\end{itemize}
\end{mmproof}
%
%
\begin{lemma}
\label{lemma:subset}
Let $F(X,Y)$ be a CNF formula and  $X'$ be a subset of variables of $X$ that are globally redundant in $F$.
Let $X''$ be a non-empty subset of $X'$. Then the variables of $X''$ are also globally redundant in $F$.
\end{lemma}
\begin{mmproof}
Assume the contrary, i.e. the variables of $X''$ are not globally redundant in $F$. Then there is an
$X''^*$-boundary point \pnt{p} where $X''^* \subseteq X''$ that is $X$-removable. Since $X''^*$ is also
a subset of $X'$, the existence of point \pnt{p} means that the variables of $X'$ are not globally redundant in $F$.
Contradiction.
\end{mmproof}
\begin{remark} 
Note that Lemma~\ref{lemma:subset} is not true for locally redundant variables. Let $F(X,Y)$ be a CNF formula
 and  $X'$ be a subset of variables of $X$ that are locally  redundant in $F$.
Let $X''$ be a non-empty subset of $X'$. Then one cannot claim that the variables of $X''$ are
locally  redundant in $F$. (However it is true that they are globally redundant in $F$.)
\end{remark}

\vspace{10pt}
For the rest of the Appendix we will use only the notion of globally redundant variables (introduced by Definition~\ref{def:glob_red_vars}).
\vspace{10pt}
%
%
\begin{definition}
\label{def:clause_cofactor}
Let $X$ be a set of Boolean variables. Let $C$ be a clause where $\V{C} \subseteq X$.
Let \Va{q} be a partial assignment to variables of $X$. Denote by \cof{C}{q} the clause that is
\begin{itemize}
\item equal to 1 (a tautologous clause) if $C$ is satisfied by \pnt{q};
\item obtained from $C$ by removing the literals falsified by \pnt{q}, if  $C$ is not satisfied by \pnt{q}.
\end{itemize}
\end{definition}
\vspace{4pt}
%
%
\begin{definition}
\label{def:local_irredundancy}
Let $F(X,Y)$ be a CNF formula and \pnt{q} be a partial assignment to variables of $X$.
Let $X'$ and $X''$ be subsets of $X$. We will say that the variables of $X''$ are \tb{locally irredundant} in 
\DDis{F}{q}{X'} if every $X''^*$-boundary point of \DDis{F}{q}{X'} where $X''^* \subseteq X''$ that is 
($X \setminus \Va{q}$)-removable in \DDis{F}{q}{X'} is  $X$-unremovable in $F$.
We will say that the variables of $X''$ \tb{are redundant in} $\boldsymbol{\mi{Dis}(F_q,X')}$ \tb{modulo local irredundancy}.
\end{definition}
%
%
\begin{remark}
The fact that variables of $X''$ are locally irredundant in \DDis{F}{q}{X'} means that the latter has an  $X''^*$-boundary point
\pnt{p} where $X''^* \subseteq X''$ that cannot be turned into a satisfying assignment in the subspace specified by \pnt{q}
(because the values of variables of \Va{q} cannot be changed). However, \pnt{p} can be transformed into a satisfying
assignment if variables of \Va{q} are allowed to be changed.  This means that \pnt{p} can be eliminated only
by an $X$-clause (implied by $F$)  but cannot be eliminated by a clause depending only
on variables of $Y$. Points like \pnt{p} can be ignored.
\end{remark}
%
%
\begin{lemma}
\label{lemma:merging_spaces}
Let $F(X,Y)$ be a CNF formula. Let \pnt{q_1} and \pnt{q_2} be partial assignments to variables of $X$
that are resolvable on variable $x$. Denote by \pnt{q} the partial assignment \Res{q_1}{q_2}{x}
(see Definition~\ref{def:res_part_assgns}).
Let $X_1$ (respectively $X_2$) be the subsets of variables of $X$ already 
proved redundant in \cof{F}{q_1} (respectively \cof{F}{q_2}).
Let the set of variables $X^*$ where $X^* = X_1 \cap X_2$ be non-empty.
Then the variables of $X^*$ are redundant in \cof{F}{q} modulo local irredundancy.
\end{lemma}
\begin{mmproof}
Assume that the variables of $X^*$ are not redundant in \cof{F}{q} and then show that this irredundancy is local.
 According to Definition~\ref{def:glob_red_vars}, irredundancy of $X^*$ means
 that there is an $X'^*$-boundary point \pnt{p} where $X'^* \subseteq X^*$
 that is ($X \setminus \Va{q}$)-removable in \cof{F}{q}.
Since \pnt{p} is an extension of \pnt{q}, it is also an extension of \pnt{q_1} or \pnt{q_2}. 
Assume for the sake of clarity that \pnt{p} is an extension of \pnt{q_1}.

The set of clauses falsified by \pnt{p} in \cof{F}{q} and \cof{F}{q_1} is specified by the set of clauses of $F$
falsified by \pnt{p}. If a clause $C$ of $F$ is satisfied by \pnt{p}, then clause \cof{C}{q}
 (see Definition~\ref{def:clause_cofactor}) is either
\begin{itemize}
\item  not in \cof{F}{q} (because is   $C$  satisfied by \pnt{q}) or
\item  in \cof{F}{q} and is satisfied by \pnt{p}.
\end{itemize}
The same applies to the relation between clause \cof{C}{q_1} and CNF formula \cof{F}{q_1}.
Let $C$ be a clause falsified by \pnt{p}. Then $C$ cannot be satisfied by \pnt{q} 
and so the clause \cof{C}{q} is in \cof{F}{q}
The same applies to \cof{C}{q_1} and \cof{F}{q_1}.

Since \pnt{p} falsifies the same clauses of $F$ in \cof{F}{q_1} and \cof{F}{q}, it is an $X'^*$-boundary point
of \cof{F}{q_1}.
 Let $P$ be the set of $2^{|X \setminus \Vaa{q_1}|}$
 points obtained from \pnt{p}
by changing  assignments to variables of $X \setminus \Va{q_1}$. Since the variables of $X^*$ are redundant
in \cof{F}{q_1}, then $P$ has to contain a point satisfying \cof{F}{q_1}. This means that 
point \pnt{p} of \cof{F}{q} can be turned into an assignment satisfying $F$ if 
the  variables that are in $\Va{q} \setminus \Va{q_1}$ are allowed to change their values. So the irredundancy
of $X^*$ in \cof{F}{q} can be only local.
\end{mmproof}
\vspace{5pt}
%
%
\begin{remark}
\label{rem:ignoring_loc_irred}
In Definition~\ref{def:d_sequent} of D-sequent \Dss{F}{X'}{q}{X''}, we did not mention local irredundancy.
However, in the rest of the Appendix we assume that the variables of $X'$ in \cof{F}{q} and those of $X''$ in \Dis{\cof{F}{q}}{X'}
may have local irredundancy. For the sake of simplicity, we do not mention this fact  with the exception of  
Lemmas~\ref{lemma:merging_red_vars} and \ref{lemma:ignoring_loc_irred}.
In particular, in Lemma~\ref{lemma:ignoring_loc_irred}, we show that D-sequents derived by \Di can only
 have local irredundancy and so the latter can be safely ignored.
\end{remark}
%
%
\begin{remark}
\label{rem:dds_impl_loc_irred}
Checking if a set of variables $X'$, where $X' \subseteq (X \setminus \Va{q})$ is irredundant in \cof{F}{q} \ti{only 
locally} is hard. For that reason \Di does not perform such a check. However, one has to introduce the notion of local irredundancy
because the latter  may appear when resolving D-sequents (see Lemma~\ref{lemma:merging_spaces}). Fortunately, 
given a D-sequent \Dss{F}{X'}{q}{X''}, one does not need to check if  irredundancy of variables $X'$ in \cof{F}{q} or $X''$ in \DDis{F}{q}{X'}
(if any) is  local. According to Lemma~\ref{lemma:ignoring_loc_irred}, this irredundancy is always local.
Eventually a D-sequent $(F,\emptyset,\emptyset) \rightarrow X$ is derived that does not have 
any local irredundancy (because the partial assignment \pnt{q} of this D-sequent is empty).
\end{remark}
\vspace{10pt}
\begin{lemma}
\label{lemma:red_vars_to_d_sequent}
Let $F(X,Y)$ be a CNF formula and \pnt{q} be a partial assignment to variables of $X$.
Let  $X^*$ where $X^* \subseteq X$ be a set of variables redundant in \cof{F}{q}.
Let sets $X'$ and $X''$ form a partition of $X^*$ i.e. $X^* = X' \cup X''$ and $X' \cap X'' = \emptyset$.
Then D-sequent \Dss{F}{X'}{q}{X''} holds.
\end{lemma}
\begin{mmproof}
Assume the contrary i.e. that the D-sequent \Dss{F}{X'}{q}{X''} does not hold.
According to Definition~\ref{def:d_sequent}, this means  that either 
\begin{itemize}
\item[A)] variables of $X'$ are not redundant in \cof{F}{q} or 
\item[B)] variables of $X''$ are not redundant in \DDis{F}{q}{X'}.
\end{itemize}

\noindent{CASE A:} This means that there exists an $X'^+$-boundary point \pnt{p} (where $X'^+ \subseteq X'$ and $\pnt{q} \leq \pnt{p}$) that
is removable in \cof{F}{q}. This implies that the variables of $X'^+$ are 
not a set of redundant variables.
On the other hand, since $X'^+ \subseteq X'$ and the variables of $X'$ are redundant, the variables of $X'^+$ are redundant too. Contradiction.

\vspace{5pt}
\noindent{CASE B:} This means that there exists an  $X''^+$-boundary point \pnt{p} (where $X''^+ \subseteq X''$ and $\pnt{q} \leq \pnt{p}$)
that is removable  in \DDis{F}{q}{X'}. Note that point \pnt{p} is an $X^{*+}$-boundary point of
 \cof{F}{q} where $X^{*+} \subseteq X^*$ (because \cof{F}{q}
consists of the clauses of \DDis{F}{q}{X'} plus some $X'$-clauses). Since the variables of $X^*$ are redundant in \cof{F}{q} the point
\pnt{p} cannot be removable. Then there is a point \pnt{p^*} obtained by flipping the variables of $X~\!\!\setminus~\!\!\Va{q}$ that satisfies \cof{F}{q}.
Point \pnt{p^*} also satisfies \DDis{F}{q}{X'}. Hence, the point \pnt{p} cannot be removable in \DDis{F}{q}{X'}. Contradiction.
\end{mmproof}
\vspace{10pt}
%
%
\begin{lemma}
\label{lemma:merging_red_vars}
Let $F(X,Y)$ be a CNF formula and \pnt{q} be a partial assignment to variables of $X$.
Let D-sequent \Dss{F}{X'}{q}{X''}  hold modulo local irredundancy. That is the variables
of $X'$ and $X''$ are redundant in \cof{F}{q} and \DDis{F}{q}{X'} respectively modulo local irredundnacy.
Then the variables of $X' \cup  X''$ are redundant in \cof{F}{q} modulo local iredundancy.
\end{lemma}
\begin{mmproof}
Denote by $X^*$ the set $X' \cup X''$.  Let \pnt{p} be a removable
$X^+$-boundary point  of \cof{F}{q} where $X^+ \subseteq X^*$. Let us consider the two possible cases:
\begin{itemize}
\item  $X^+ \subseteq X'$ (and so $X^+ \cap X'' = \emptyset$).
Since \pnt{p} is removable, the variables of $X'$ are irredundant in \cof{F}{q}. Since this irredundancy can only be local
one can turn \pnt{p} into an assignment satisfying $F$.
 This means that the irredundancy of variables $X^*$ in $F$ due to point \pnt{p} is local.
\item  $X^+ \not\subseteq X'$ (and so $X^+ \cap X'' \neq \emptyset$). 
 Then \pnt{p} is an $X''^+$-boundary
 point of \DDis{F}{q}{X'} where $X''^+ = X^+ \cap X''$. 
Indeed, for every variable $x$ of $X^+$ there has to be a clause  $C$ of \cof{F}{q} 
falsified by \pnt{p} such that $\V{C}~\cap X^+~=~\s{x}$.
 Otherwise, $x$ can be removed from $X^+$, which contradicts the assumption that \pnt{p} is an $X^+$-boundary point.
 This means that for every variable $x$ of $X''^+$ there is a clause $C$ falsified by \pnt{p} such 
that $\V{C}~\cap X''^+~=~\s{x}$. 

\vspace{10pt}
Let $P$ denote the set of all $2^{|X \setminus (\mi{Vars}(\boldsymbol{q}) \cup X')|}$ points obtained from \pnt{p} by flipping 
values of variables of $X \setminus (\Va{q} \cup X')$.
Let us consider the following two possibilities.
     \begin{itemize}
        \item Every point of $P$ falsifies \DDis{F}{q}{X'}. This means that the point \pnt{p}
           is a removable $X''^+$-
               boundary point of \DDis{F}{q}{X'}. Hence the variables of $X''$ are  irredundant in \DDis{F}{q}{X'}.
           Since this irredundancy is local, point \pnt{p} can be turned into an assignment satisfying  $F$ by changing
           values of variables of $X$.  Hence the irredundancy of $X^*$ in $F$ due to point \pnt{p} is local.
         \item A point \pnt{d} of $P$ satisfies \DDis{F}{q}{X'}. Let us consider the following two cases.
            \begin{itemize}
              \item \pnt{d} satisfies \cof{F}{q}. This contradicts the fact that \pnt{p} is
 a removable $X^+$-boundary point of \cof{F}{q}.
                      (By flipping variables of $X \setminus \Va{q}$
                       one can  obtain a point satisfying \cof{F}{q}.)

              \item \pnt{d} falsifies some clauses of \cof{F}{q}. Since \cof{F}{q} and \DDis{F}{q}{X'}
 are different only in $X'$-clauses,
                            \pnt{d} is an $X'^*$-boundary point of \cof{F}{q} where $X'^* \subseteq X'$. Since
 \pnt{p} is a removable $X^+$-boundary point
                            of \cof{F}{q}, \pnt{d} is a removable  $X'^*$-boundary point of \cof{F}{q}.
            So the variables of $X'$ are irredundant in \cof{F}{q}. Since this irredundancy is local,
the point \pnt{d} can be turned into an assignment satisfying $F$ by changing the values of $X$. Then,
the same is true for point \pnt{p}. So the irredundancy of $X^*$  in $F$ due to point \pnt{p} is local.
            \end{itemize}
     \end{itemize}
\end{itemize}
\end{mmproof}
%
%
\vspace{10pt}
\begin{proposition}
Let $F(X,Y)$ be a CNF formula. Let D-sequents $S_1$ and $S_2$ be equal to  \Dss{F}{X_1}{q_1}{X'} and
\Dss{F}{X_2}{q_2}{X'} respectively. Let \pnt{q_1} and \pnt{q_2}
be resolvable on variable $x$.  Denote by \pnt{q} the partial assignment \Res{q_1}{q_2}{x}
and by $X^*$ the set $X_1 \cap X_2$.
Then, if $S_1$ and $S_2$ hold, the D-sequent $S$ equal to \Dss{F}{X^*}{q}{X'} holds  too.
\end{proposition}
\begin{mmproof}
Lemma~\ref{lemma:merging_red_vars} implies  that the variables of $X_1 \cup X'$ and  $X_2 \cup X'$ 
are redundant in \cof{F}{q_1} and \cof{F}{q_2} respectively. From Lemma~\ref{lemma:merging_spaces}, one concludes
that the variables of  the set $X'' =  (X_1 \cup X') \cap (X_2 \cup X')$ are redundant in \cof{F}{q}.
From Definition~\ref{def:d_sequent} it follows that $X_1 \cap X' = X_2 \cap X' = \emptyset$.
So $X'' = X^* \cup X'$ where $X^* \cap X' = \emptyset$.
Then, from Lemma~\ref{lemma:red_vars_to_d_sequent}, it follows that
the D-sequent   \Dss{F}{X^*}{q}{X'} holds.
\end{mmproof}

\subsection*{SUBSECTION: Derivation of a D-sequent}
\begin{proposition}
Let $F(X,Y)$ be a CNF formula and \pnt{q} be a partial assignment to variables of $X$. 
Let \Su{X}{red} be the variables proved  redundant in  \cof{F}{q}.
Let $x$ be the only variable of $X$ that is not in $\Va{q} \cup \Su{X}{red}$.
Let D-sequent \DDs{F}{\Su{X}{red}}{q}{x} hold.
Then D-sequent \DDs{F}{\Su{X'}{red}}{g}{x} holds where   \pnt{g} and $\Su{X'}{red}$ are defined as follows.
Partial assignment \pnt{g} to variables of $X$ satisfies   
the two  conditions below (implying that $\pnt{g} \leq \pnt{q}$):
\begin{enumerate}
\item Let $C$ be a \s{x}-clause of $F$ that is not in \DDis{F}{q}{\Su{X}{red}}. Then either
  \begin{itemize}
   \item \pnt{g} contains an assignment satisfying $C$ or 
    \item  D-sequent  \DDs{F}{\Su{X^*}{red}}{g^*}{x^*} holds where
      $\pnt{g^*}~\leq~\pnt{g}$, $\Su{X^*}{red} \subset \Su{X}{red}$, $x^* \in (\Su{X}{red} \cap \V{C})$.
  \end{itemize}
\item Let \pnt{p_1} be a point such that  $\pnt{q} \leq \pnt{p_1}$. Let \pnt{p_1} 
falsify a clause of $F$ with literal $x$. Let \pnt{p_2} be obtained from \pnt{p_1} by 
flipping the value of $x$
and falsify a clause of $F$ with literal $\overline{x}$. Then there is a non-\s{x}-clause $C$ of $F$ falsified
by \pnt{p_1} and \pnt{p_2} such that  $(\V{C} \cap X) \subseteq \Va{g}$.
\end{enumerate}
The set \Su{X'}{red} consists of all the variables already proved redundant in \cof{F}{g}. That is every  redundant
variable $x^*$ of \Su{X}{red} with D-sequent \DDs{F}{\Su{X^*}{red}}{g^*}{x^*} such 
that $\pnt{g^*}~\leq~\pnt{g}$, $\Su{X^*}{red} \subset \Su{X}{red}$
is in \Su{X'}{red}.
\end{proposition}
\begin{mmproof}
Assume the contrary i.e. D-sequent   \DDs{F}{\Su{X'}{red}}{g}{x} does not hold, and so
 variable $x$ is not redundant in 
\DDis{F}{g}{\Su{X'}{red}}. Hence there is a point \pnt{p}, $\pnt{g} \leq \pnt{p}$ that is a removable \s{x}-boundary point
of  \DDis{F}{g}{\Su{X'}{red}}.

Let $C$ be an \s{x}-clause of $F$. Note that  \DDis{F}{g}{\Su{X'}{red}} cannot contain the clause \cof{C}{g} if 
the clause \cof{C}{q} is not in \DDis{F}{q}{\Su{X}{red}}.
If \cof{C}{q} is not in  \DDis{F}{q}{\Su{X}{red}}, then \pnt{g} either satisfies $C$ or $C$ 
contains a variable of \Su{X}{red} that is also in \Su{X'}{red} (and hence \cof{C}{g} contains
a redundant variable and so is not in \DDis{F}{g}{\Su{X'}{red}}).

So, for \pnt{p} to be an \s{x}-boundary point of \cof{F}{g}, there has to be 
\s{x}-clauses  $A$ and $B$ of $F$ such that 
\begin{itemize}
\item  they are not satisfied by \pnt{g} and do not contain variables of \Su{X'}{red}
       (so the  clauses \cof{A}{g} and \cof{B}{g} are in \DDis{F}{g}{\Su{X'}{red}})
\item  $A$ is  falsified by \pnt{p} and $B$ is falsified by the point 
obtained from \pnt{p} by flipping the value of $x$.
\end{itemize}
 Let point \pnt{p_1} 
 be obtained from \pnt{p} by 
flipping assignments to the variables of $\Va{q} \setminus \Va{g}$ that disagree with \pnt{q}.
 By construction $\pnt{g} \leq \pnt{p_1}$ and
$\pnt{q} \leq \pnt{p_1}$. Let \pnt{p_2} be the point obtained from \pnt{p_1} by flipping the
value of $x$. Since $x$ is not assigned in \pnt{q} (and hence is not assigned in \pnt{g}),
$\pnt{g} \leq \pnt{p_2}$ and $\pnt{q} \leq \pnt{p_2}$.
Then  \cof{A}{q} and \cof{B}{q}  are also in \cof{F}{q}. As we mentioned above  $A$ and $B$ cannot
contain variables of $\cof{X}{red}$ (otherwise they could not be in \cof{F}{g}).
So $A$ and $B$ are also in \DDis{F}{q}{\Su{X}{red}}.

Note that clause $A$ is falsified by \pnt{p_1}. Assume the contrary, i.e. that  $A$ is satisfied by \pnt{p_1}.
Then the fact that  \pnt{p} and \pnt{p_1} are different only in assignments to \pnt{q} and that \pnt{p}
falsifies $A$ implies that $q$ satisfies $A$. But then by construction, \pnt{g} has to satisfy $A$ and
we have contradiction. Since $B$ is also an \s{x}-clause as $A$, one can use the same reasoning to show that 
\pnt{p_2} falsifies $B$.

Since  \pnt{p_1} and \pnt{p_2} falsify \s{x}-clauses $A$ and $B$ and $\pnt{p_1},\pnt{p_2} \leq \pnt{q}$
one can apply Condition 2 of the proposition
at hand. That is  there must be a clause $C$ falsified by \pnt{p_1} and \pnt{p_2} such that 
\pnt{g}  contains all the assignments of \pnt{q} that falsify literals of $C$. This means that
 $C$ is not satisfied by \pnt{g}.
Besides, since due to Condition 2  every variable of $\V{C} \cap X$ is in \Va{g},
every variable of \cof{C}{g} is in $Y$. Hence a variable of \cof{C}{g} cannot be redundant.
This means that \cof{C}{g} is   in  \DDis{F}{g}{\Su{X'}{red}}. Since \pnt{p} and \pnt{p_1} have
identical assignments to the variables of $Y$, then \pnt{p} falsifies  \cof{C}{g} too. So
\pnt{p} cannot be an \s{x}-boundary point of  \DDis{F}{g}{\Su{X'}{red}}. Contradiction.
\end{mmproof}

%
%
\section*{Proofs of Section~\ref{sec:dds_impl}}

%
%
\begin{lemma}
\label{lemma:ignoring_loc_irred}
Let \Dss{F}{X'}{g}{X''} be a D-sequent derived by \ti{DDS\_impl} and \pnt{q} be the partial assignment
when this D-sequent is derived.  Let variables of $X'$ be irredundant in 
\cof{F}{g} or variables of $X''$ be irredundant in 
\DDis{F}{g}{X'}. Then this irredundancy is only local. (See Definition~\ref{def:local_irredundancy}
 and Remarks~\ref{rem:ignoring_loc_irred}
 and \ref{rem:dds_impl_loc_irred}.)
\end{lemma}
\begin{mmproof}
We carry out the proof by induction in the number of D-sequents. The base step is that the statement holds 
for an empty set of D-sequents,
which is vacuously true. The inductive step is to show that  the fact that the statement holds for D-sequents
 $S_1, \ldots, S_n$ implies
that it is true for $S_{n+1}$. Let us consider all possible cases. 
\begin{itemize}
\item $S_{n+1}$ is a  D-sequent \DDs{F}{X'}{g}{x} for a monotone variable $x$ of \DDis{F}{g}{X'}
 where $x \in (X \setminus (\Va{q} \cup X')$.
      Since formula \DDis{F}{g}{X'} cannot have removable \s{x}-boundary points 
(see Proposition~\ref{prop:mon_var_red}), variable $x$
      cannot be irredundant in \DDis{F}{g}{X'}.  The variables of $X'$ may be irredundant in \cof{F}{g}.
 However, this irredundancy can be
      only local. Indeed,  using Lemma~\ref{lemma:merging_red_vars} and the 
induction hypothesis one can show that  variables proved redundant for \cof{F}{g}
according to the relevant D-sequents of the set  $\{S_1,\ldots,S_n\}$ are indeed redundant in \cof{F}{g} modulo local
irredundancy.
\item $S_{n+1}$ is a D-sequent \Dss{F}{\emptyset}{g}{X'} derived due to appearance 
of an empty clause $C$ in \cof{F}{g}. Here 
\pnt{g} is the minimum subset of assignments of \pnt{q} falsifying $C$.
 In this case,  \cof{F}{g}
has no boundary points and hence the set $X'$ of unassigned variables of  \cof{F}{g} cannot be irredundant.
\item $S_{n+1}$ is a D-sequent \DDs{F}{X'}{g}{x} derived after making the only unassigned variable $x$ of 
 \DDis{F}{q}{\Su{X}{red}} redundant by adding resolvents on variable $x$. (As usual, \Su{X}{red} denotes the
 set of redundant variables already
proved redundant in \cof{F}{q}.)
In this case, every removable \s{x}-boundary point of  \DDis{F}{q}{\Su{X}{red}} is eliminated and so the 
latter cannot be irredundant in $x$.
Due to Proposition~\ref{prop:d_seq_derivation}, the same applies to \DDis{F}{g}{X'}. To show that 
 irredundancy of variables of $X'$ in \cof{F}{g}
can be only local one can use the same reasoning as in the case when $x$ is a monotone variable.
\item $S_{n+1}$ is obtained by resolving D-sequents $S_i$ and $S_j$ where $1 \leq i,j \leq n$
 and $i \neq j$. Let $S_i$,$S_j$ and $S_{n+1}$ be equal 
to \Dss{F}{X_i}{q_i}{X''}, \Dss{F}{X_j}{q_j}{X''} and \Dss{F}{X'}{g}{X''} respectively where $X' = X_i \cap X_j$
and \pnt{g} is obtained by resolving \pnt{q_i} and \pnt{q_j} (see Definition~\ref{def:res_part_assgns}).

\vspace{5pt}
Let us first show that irredundancy of $X''$ in \DDis{F}{g}{X'} can only be local. 
Let \pnt{p} be a removable  $X''^*$-boundary point of \DDis{F}{g}{X'} where $X''^* \subseteq X''$.

 Then either $\pnt{q_i} \leq \pnt{p}$
or $\pnt{q_j} \leq \pnt{p}$.
Assume for the sake of clarity that  $\pnt{q_i} \leq \pnt{p}$. Consider the following two cases.
   \begin{itemize}
    \item \pnt{p} is not removable in \DDis{F}{q_i}{X_i}. Then the irredundancy of $X''$ in \DDis{F}{g}{X'}
 due to point \pnt{p} is local. (A point satisfying \DDis{F}{q_i}{X_i} can be obtained from \pnt{p} by changing
 values of some variables from
 $X \setminus (X_i \cup \Va{q_i})$. The same point satisfies  \DDis{F}{g}{X'} because $\pnt{g} \leq \pnt{q_i}$
 and $X' \subseteq X_i$.)
    \item \pnt{p} is also removable in  \DDis{F}{q_i}{X_i}. This means that the variables of $X'$ are 
irredundant in \DDis{F}{q_i}{X_i}.
          By the induction hypothesis, this irredundancy is  local. Then one can turn \pnt{p} into a
 satisfying assignment of $F$ by changing
          assignments to variables of $X$. Hence the  irredundancy of $X''$ in \DDis{F}{g}{X'}
 due to point \pnt{p} is also local.
    \end{itemize}
  \vspace{5pt}
 Now, let us show that irredundancy of $X'$ in \cof{F}{g} can only be local.
Let \pnt{p} be a removable  $X'^*$-boundary point of \cof{F}{g} where $X'^* \subseteq X'$.
Again, assume for the sake of clarity that  $\pnt{q_i} \leq \pnt{p}$. Consider the following two cases.
     \begin{itemize}
    \item \pnt{p} is not removable in \cof{F}{q_i}. Then the irredundancy of $X'$ in \cof{F}{q} due 
to point \pnt{p} is local. (A point satisfying  \cof{F}{q_i} can be obtained by from \pnt{p} 
by changing values of some variables from $X \setminus \Va{q_i}$. The same point satisfies  \cof{F}{g}
because $\pnt{g} \leq \pnt{q_i}$.)
    \item \pnt{p} is also removable in  \cof{F}{q_i}. This means that the variables of $X'$ 
(and hence the variables of $X_i$) 
     are irredundant in \cof{F}{q_i}.
          By the induction hypothesis, this irredundancy is  local. Then one can turn \pnt{p} 
into a satisfying assignment of $F$ by changing
          assignments to variables of $X$. Hence the  irredundancy of $X'$ in \cof{F}{q}
 due to point \pnt{p} is also local.
     \end{itemize}
\end{itemize}
\end{mmproof}
\begin{remark}
Note that correctness of the final D-sequent $(F,\emptyset,\emptyset) \rightarrow X$ modulo local irredundancy 
implies that  the variables of $X$ are redundant in $F$.
In this case, there is no difference between just redundancy and redundancy modulo local irredundancy
 because \pnt{q} is empty.
(So the value of any variable of $X$ can be changed when checking if a boundary point is removable.)
\end{remark}
%
%
\begin{lemma}
\label{lemma:final_d_sequent}
Let $F(X,Y)$ be a CNF formula and $X= \{x_1,\ldots,x_k\}$.
Let $S_1,\ldots,S_k$ be D-sequents where $S_i$ is the D-sequent $\emptyset \rightarrow \s{x_i}$.
Assume that $S_1$ holds for the formula $F$, $S_2$ holds for the formula \Dis{F}{\s{x_1}},
$\ldots$,$S_{k}$ holds for the formula \Dis{F}{\{x_1,\ldots,x_{k-1}\}}.
(To simplify the notation we assume that  D-sequents $S_i$ have been derived in the order they are numbered).
Then the variables of $X$ are redundant in $F(X,Y)$.
\end{lemma}
\begin{mmproof}
Since $S_1$ holds, due to Proposition~\ref{prop:red_vars}, the formula $\exists{X}.F$
is equivalent to $\exists(X \setminus \s{x_1})$.\Dis{F}{\s{x_1}}. Since $S_2$ holds for
\Dis{F}{\s{x_1}} one can apply  Proposition~\ref{prop:red_vars} again to show that
$\exists(X \setminus \s{x_1})$.\Dis{F}{\s{x_1}} is equivalent to 
$\exists(X \setminus \{x_1,x_2\})$.\Dis{F}{\{x_1,x_2\}} and hence the latter is equivalent to
$\exists{X}.F$. By applying Proposition$\,$\ref{prop:red_vars}~~  $k\!\!-\!\!2$ more times one shows
that $\exists{X}.F$ is equivalent to \Dis{F}{X}. According to Corollary~\ref{corol:red_vars}, this means that
the variables of $X$ are redundant in $F(X,Y)$. 
\end{mmproof}
%
%
\begin{proposition}
\Di is sound and complete.
\end{proposition}
\begin{mmproof}
First, we show that \Di is complete. \Di builds a binary search tree and
visits every node of this tree at most three times (when starting the left branch, when backtracking
to start the right branch, when backtracking from the right branch). 
So \Di is complete.

Now we prove that \Di is sound. The idea of the proof is to show that all D-sequents derived by 
\Di are correct. By definition, \Di eventually derives correct D-sequents $\emptyset \rightarrow \s{x}$ for 
every variable of $X$. From Lemma~\ref{lemma:final_d_sequent} it follows that this is equivalent
to derivation of the  correct D-sequent $\emptyset \rightarrow X$.

We prove the correctness of D-sequents derived by \Di by induction. 
The base statement is that the D-sequents of an empty set are correct (which is vacuously true).
The induction step  is that to show that if first $n$ D-sequents are correct, then next D-sequent $S$ 
is correct too. Let us consider the following alternatives.
\begin{itemize}
\item $S$ is a D-sequent built for  a monotone variable of \Fcurr\!. 
The correctness of $S$ follows from  Proposition~\ref{prop:d_seq_derivation}
and the induction hypothesis (that the D-sequents derived before are correct).
\item $S$ is the D-sequent specified by a  locally empty clause. In this case, $S$ is trivially true.
\item  $S$ is a D-sequent derived by \Di in the BPE state for variable $x$ 
after  eliminating \s{x}-removable \s{x}-boundary points  of \DDis{F}{q}{\Su{X}{red}}.
The correctness of $S$ follows form Proposition~\ref{prop:d_seq_derivation} and the induction hypothesis. 
\item 
 $S$ is obtained by resolving  two existing D-sequents. The correctness of $S$ follows from 
 Proposition~\ref{prop:res_rule} and the induction hypothesis.

\end{itemize}
\end{mmproof}
\vspace{10pt}
%
%
\section*{Proofs of Section~\ref{sec:compos}}
%
%
\begin{definition}
Let \ti{Proof} be a resolution proof that a CNF formula $H$ is unsatisfiable.
 Let \Su{G}{proof} be the resolution
graph specified by \ti{Proof}. (The sources of \Su{G}{proof} correspond to  
clauses of $H$. Every non-source node of \Su{G}{proof} 
corresponds to a resolvent of \ti{Proof}. The sink of \Su{G}{proof} is an empty clause.
Every non-source node of \Su{G}{proof} has two incoming edges connecting this note to the
nodes corresponding to the parent clauses.)
 We will call \ti{Proof} \tb{irredundant},
if for every node of \Su{G}{proof} there is a path leading from this node to the sink.
\end{definition}
%
%
\begin{lemma}
\label{lemma:comp_proof}
Let  $F(X,Y)$ be equal to $F_1(X_1,Y_1) \wedge \ldots \wedge F_k(X_k,Y_k)$ where
$(X_i \cup Y_i) \cap (X_j \cup Y_j) = \emptyset$, $i \neq j$. Let $F$ be satisfiable.
Let $F$ have no \s{x}-removable \s{x}-boundary points where $x \in X_i$ and \ti{Proof} be 
a resolution proof of that fact built by \Di\!\!. 
Then \ti{Proof} does not contain clauses of $F_j$,$j \neq i$ (that is no clause
of $F_j$ is used as a parent clause in a resolution of \ti{Proof}).
\end{lemma}
\begin{mmproof}
\Di concludes that all \s{x}-removable \s{x}-boundary points have been eliminated if the CNF formula $H$ 
described in Subsection~\ref{subsec:merging_branches}  is unsatisfiable.
 $H$ consists of  clauses of the current formula \DDis{F}{q}{\Su{X}{red}} and the clauses of 
CNF formula \Su{H}{dir}. 
\Di builds an irredundant resolution proof that $H$ is unsatisfiable. 
(Making \ti{Proof} irredundant is performed by function \ti{optimize} of Figure~\ref{fig:bpe}.)

Since formula $F$ is the conjunction of independent subformulas, clauses of $F_i$ and $F_j$, $j \neq i$
 cannot be resolved with each other.
The same applies to \ti{resolvents} of clauses of $F_i$ and $F_j$
and to resolvents of clauses of $F_i \wedge \Su{H}{dir}$ and $F_j$. (By construction~\cite{HVC},
 \Su{H}{dir} may have only variables
of \s{x}-clauses of $F$  and  some new variables i.e. ones that are not present in $F$. Since
$x \in X_i$, this means  that the variables of \Su{H}{dir} can only overlap with those of $F_i$.)
Therefore, an irredundant proof of unsatisfiability of $H$ 
has to contain only clauses of either formula $F_j$, $j\neq i$ or  formula
 $F_i \wedge \Su{H}{dir}$.
Formula $F$ is satisfiable, hence every subformula $F_j$, $j=1,\ldots,k$ is satisfiable too. So, a proof cannot 
consists solely of
clauses of $F_j$,$j \neq i$.  This means that \ti{Proof} employs only clauses of 
 $F_i \wedge \Su{H}{dir}$ (and their resolvents).

\end{mmproof}
\vspace{10pt}
%
%
\begin{proposition}
\Di is compositional regardless of  how branching variables are chosen. 
\end{proposition}
\begin{mmproof}
The main idea of the proof  is to show that every D-sequent generated by  \Di  has the form \Dds{g}{X'} 
where $\Va{g} \subseteq X_i$  and $X' \subseteq X$.
We will call such a D-sequent \tb{limited to} \pnt{F_i}. 
Let us carry on the proof by induction. Assume that the
D-sequents generated so far are limited to $F_i$  and show that this holds for the next D-sequent $S$.
Since one cannot resolve clauses of $F_i$ and $F_j$, $i \neq j$, if $S$ is specified by
a clause that is locally empty, $S$ is limited to $F_i$.  

Let $S$ be a D-sequent generated 
for a monotone variable $x \in X_i$. According to Remark~\ref{rem:mon_var}, only Condition~\ref{enum:x_clauses}
contributes to forming \pnt{g}. In this case, \Va{g} consists of 
\begin{enumerate}
\item variables of \s{x}-clauses of $F$ and
\item variables of \Va{g^*} of D-sequents  $\pnt{g^*} \rightarrow \s{x^*}$
 showing redundancy of variables $x^*$ of 
\s{x}-clauses of $F$.
\end{enumerate}
Every \s{x}-clause of $F$ is either a clause of the original formula $F_i$ or its resolvent. So the variables
that are in \pnt{g} due to the first condition above are in $X_i$. By the induction hypothesis, the variables
of \Va{g^*} are also in $X_i$.

Let $S$ be obtained after eliminating \s{x}-removable \s{x}-boundary points where $x \in X_i$ 
(see Subsection~\ref{subsec:merging_branches}). Denote by \pnt{g_1} and \pnt{g_2} the two parts of \pnt{g}
specified by Condition~\ref{enum:x_clauses} and~\ref{enum:non_x_clauses} of Proposition~\ref{prop:d_seq_derivation}.
(Assignment \pnt{g} is the union of assignments \pnt{g_1} and \pnt{g_2}.) The variables of \Va{g_1} are
in $X_1$ for the same reasons as in the case of monotone variables.

To generate \pnt{g_2}, \Di uses   proof
\ti{Proof} that formula $H$ built from clauses of $F$ and $\Su{H}{dir}$  
(see Subsection~\ref{subsec:merging_branches}) is unsatisfiable.
As we showed in Lemma~\ref{lemma:comp_proof}, \ti{Proof} employs only clauses 
of $F_i \wedge \Su{H}{dir}$ and their resolvents. Only clauses of formula $F$ are
taken into account when forming \pnt{g_2} in Proposition~\ref{prop:d_seq_derivation}
 (i.e. clauses of \Su{H}{dir} do  not affect \pnt{g_2}).
Since the only clauses of $F$ used   in \ti{Proof} are those of $F_i$, then  $\Va{g_2} \subseteq X_i$.

Finally, if $S$ is obtained by resolving two D-sequents limited to $F_i$, it is also limited to $F_i$
(see Definition~\ref{def:res_rule}). 
\end{mmproof}

\end{document}